\documentclass[a4paper,11pt]{article}
\pdfoutput=1 
\usepackage{tabls}
\usepackage{color}
\usepackage{bbold}
\usepackage{jcappub} 
\usepackage{comment}
\usepackage[T1]{fontenc} 
\usepackage{dsfont}
\title{All-sky reconstruction of the primordial scalar potential from WMAP temperature data}
\author[a,b,1]{Sebastian Dorn,\note{Corresponding author.}}
\author[a,b]{Maksim Greiner,}
\author[a,b]{and Torsten A.\ En{\ss}lin}
\affiliation[a]{Max-Planck-Institut f\"ur Astrophysik,\\ Karl-Schwarzschild-Str.~1, D-85748 Garching, Germany}
\affiliation[b]{Ludwigs-Maximilians-Universit\"at M\"unchen,\\ Geschwister-Scholl-Platz 1, D-80539 Munich, Germany}
\emailAdd{sdorn@mpa-garching.mpg.de}\emailAdd{maksim@mpa-garching.mpg.de}\emailAdd{ensslin@mpa-garching.mpg.de }

\abstract{An essential quantity required to understand the physics of the early Universe, in particular the inflationary epoch,
is the primordial scalar potential $\Phi$ and its statistics. We present for the first time an all-sky reconstruction of $\Phi$ with corresponding 
$1\sigma$-uncertainty from WMAP's cosmic microwave background (CMB)
temperature data -- a map of the very early Universe right after the inflationary epoch. This has been achieved by applying
a Bayesian inference method that separates the whole inverse problem of the reconstruction into many independent ones, each of them solved
by an optimal linear filter (Wiener filter). In this way, the three-dimensional potential $\Phi$ gets reconstructed slice by slice resulting in a thick
shell of nested spheres around the comoving distance to the last scattering surface. Each slice represents the primordial scalar 
potential $\Phi$ projected onto a sphere with corresponding distance. Furthermore, we present an advanced method for inferring $\Phi$ and its
power spectrum simultaneously from data, but argue that applying it requires polarization data with high signal-to-noise levels not available yet.
Future CMB data should improve results significantly, as polarization data will fill the present $\ell-$blind gaps of the reconstruction.}

\begin{document}

\excludecomment{commenta}
\maketitle
\flushbottom
\section{Introduction \& motivation}
The cosmic microwave background radiation (CMB) is presently one of the most informative data sets for cosmologists to study the physics of the early Universe.
Of actual interest is in particular the verification of the existence  of an inflationary phase of the Universe and investigations of the physical properties of
the involved inflaton field(s). An essential quantity is thereby the
primordial adiabatic scalar potential $\Phi$. Its statistic, especially the two-point function, was determined during inflation, when the quantum
fluctuations of the inflationary field were frozen during their exit of the Hubble horizon. This statistic is conserved on super-horizon scales during the epoch of
reheating until the individual perturbed modes re-enter the horizon.
Therefore, significant information on the inflationary phase is encoded
in the observable quantity $\Phi$. 
The processes translating the initial modes after their horizon re-entry into the observed CMB fluctuations are described by the so-called radiation transfer functions,
see Refs.~\cite{2009arXiv0907.5424B,2010AdAst2010E..71Y}. As a consequence, many inference methods aim at constraining parameters of the early Universe
involve $\Phi$ or their statistics. Therefore the CMB fluctuations provide a highly processed view on the primordial scalar potential.
In this work, we attempt, however, their direct reconstruction and visualization via Bayesian inference. Once they are reconstructed a direct investigation of their
statistics is possible, e.g., the inference of the primordial power spectrum, their connection to large scale structure \cite{2013MNRAS.432..894J},
or primordial magnetic fields \cite{2014JCAP...07..012N,2005PhRvD..71d3502M}.

The Planck observation,  Ref.~\cite{2013arXiv1303.5084P}, of the almost homogeneous and isotropic CMB have shown that the statistical deviations
from Gaussianity of the primordial modes/perturbations
are still consistent with zero. Therefore, the two-point correlation function of $\Phi$ seems to describe nearly fully the statistics of the
early Universe up to high accuracy.
This fact simplifies the inference of these modes significantly (see, e.g., Ref.~\cite{paper1,2014JCAP...06..048D}), and enables a well justified all-sky reconstruction
of the primordial scalar potential from real data.

This work is organized as follows. In Sec.~\ref{sec:inf} we present a Bayesian inference approach to reconstruct the
primordial scalar potential. This method, initially proposed by Ref.~\cite{2010AdAst2010E..71Y}, requires the knowledge of the primordial power spectrum. We show
further how $\Phi$ and its spectrum can be inferred (unparametrized) even without such an a priori knowledge or assumption.
In Sec.~\ref{sec:recon}, we reconstruct the primordial
scalar potential with corresponding $1\sigma$-uncertainty from WMAP temperature data \cite{2013ApJS..208...20B} and partially its initial power spectrum.
In Sec.~\ref{sec:conclusion}, we summarize our findings. Exact
derivations of all used reconstruction methods can be found in appendices~\ref{app:response_der}-\ref{app:noise}.

\section{Inference approach}\label{sec:inf}
We derive the inference methods within the framework of information field theory (IFT) \cite{2009PhRvD..80j5005E},
where $\Phi$ is considered to be a physical scalar field, defined over the Riemannian manifold $\mathds{R}^3$. Since there is no solid evidence that $\Phi$ is non-Gaussian,
we assume its statistics to be Gaussian with a covariance matrix determined by its power spectrum\footnote{Here we assume that $\Phi$
is also statistically homogeneous and isotropic.}, i.e.,
\begin{equation} 
\Phi \hookleftarrow \mathcal{G}(\Phi, P^\Phi)~~\mathrm{with}~~P^\Phi(k,q) \equiv {\left\langle \Phi\Phi^\dag \right\rangle}_{(\Phi)}=(2\pi)^3\delta(k-q) P^\Phi(k).
\end{equation}
Thereby we introduced the notation
\begin{equation}
\mathcal{G}(a, A) \equiv \frac{1}{\sqrt{|2\pi A|}}\exp\left(-\frac{1}{2}a^\dag {A}^{-1}a \right)~~\mathrm{and}~~
\left\langle{~.~}\right\rangle_{(a)} \equiv \int \mathcal{D}a{~.~}\mathcal{G}(a, A),
\end{equation}
with corresponding inner product 
\begin{equation}
a^\dag b \equiv \int_{\mathds{R}^3}d^{3}x ~a^*(x)b(x)
\end{equation}
for the fields $a,~b$. Here, $\dag$ denotes a transposition, $t$, and complex conjugation, $*$.
The CMB data, on the other hand, are of discrete nature, i.e., $d\equiv \left(d_1,\dots,d_n \right)^t \in \mathds{R}^n,~n\in \mathds{N}$.

\subsection{Temperature only}
To set up a Bayesian inference scheme for the primordial scalar potential $\Phi$ we have to know how the data $d$ are related to $\Phi$.
In the case of the data being the WMAP CMB temperature map this relation is well known, given by \cite{2005ApJ...634...14K}
\begin{equation}\label{data}
\begin{split}
 d_{\ell m} \equiv&~ \left(R \Phi\right)_{\ell m} + n_{\ell m}\\
 	=&~ M_{\ell m \atop \ell' m'} B_{\ell'} ~\frac{2}{\pi}\int dk ~k^2 \int dr ~r^2 \Phi_{\ell' m'}(r) g^T_{\ell'}(k)j_{\ell'}(kr) + n_{\ell m},
\end{split} 								
\end{equation}
where $g^T_\ell(k)$ denotes the adiabatic radiation transfer function of temperature, $j_\ell (kr)$ the spherical Bessel function, $n\in\mathds{R}^n$ the
additive Gaussian noise, and $B_\ell$ the beam transfer function of the WMAP satellite. Repeated indices are implicitly summed over unless they are free on both sides of the equation.
We assume the noise to be uncorrelated to $\Phi$. 
The operator $R$, which transforms $\Phi$ into the CMB temperature map,
is assumed to be linear consisting of an integration in Fourier space as well as over the radial (comoving distance) coordinate plus the instrument's beam
convolution and a foreground mask, $M$. Since there is currently no hint for isocurvature modes \cite{2014A&A...571A..22P} we exclude them from all calculations.

The next logical step, the construction of an optimal\footnote{Optimal with respect to the $\mathcal{L}^2-$error norm.} linear filter within the framework of IFT,
e.g. the Wiener filter \cite{wiener1964time} (see, e.g., Ref.~\cite{2009PhRvD..80j5005E}), is straightforward. Given the actual, very high resolution
of current CMB data sets this, however, turns out to be extremely expensive. 

Fortunately, there is a way to split this single computation of reconstructing the primordial scalar potential into multiple.
Instead of reconstructing the three-dimensional $\Phi$ in a single blow, one can reconstruct it spherically slice by slice, each slice corresponding
to a specific radial coordinate starting from $r=0$ to beyond the surface of last scattering (LSS), $r_\mathrm{LSS}$. 
To understand this procedure we want to recall the definition of the response stated in Ref.~\cite{2009PhRvD..80j5005E}, where $R$ is the part of the data 
which correlates with the signal, $R\Phi=\left\langle d \right\rangle_{(d|\Phi)}$. It is straightforward to show that this is equivalent to 
\begin{equation}
\label{response_3d_1}
R \equiv \left\langle d \Phi^\dag \right\rangle_{(\Phi,d)} \left\langle \Phi \Phi^\dag \right\rangle^{-1}_{(\Phi,d)}.
\end{equation}
To obtain the response acting on a sphere with corresponding comoving distance $r$ it can now also be defined as the expectation value of the data given $\Phi$ restricted to a sphere instead of over
the three-dimensional regular space, i.e., $R^{(2)}\Phi(r=\mathrm{const.})=\left\langle d \right\rangle_{(d|\Phi(r=\mathrm{const.}))}$. The exact derivation of this modification can be found
in App.~\ref{app:response_der} and yields
\begin{equation}
\begin{split}
\label{R2}
 R_{\ell m \atop \ell' m'}^{(2)}(r) =&~M_{\ell m \atop \ell'' m''} B_{\ell''} \frac{\int dk~ k^2 P^\Phi(k) j_{\ell''}(kr) g^T_{\ell''}(k)}
 {\int dk ~k^2 P^\Phi(k) j^2_{\ell''}(kr)}\delta_{\ell'' \ell'}\delta_{m'' m'}\\
 \equiv &~M_{\ell m \atop \ell'' m''} B_{\ell''} R_{\ell''}\delta_{\ell'' \ell'}\delta_{m'' m'},
\end{split}
\end{equation}
with 
superscript ``$(2)$'' indicating that this response acts on the (two-dimensional)
sphere $\Phi_{\ell m}(r=\mathrm{const.})$. Initially, we assume $P^\Phi$ to be known (see Sec.~\ref{sec:crit} if not), i.e. that it 
is determined via the primordial power spectrum of comoving curvature perturbations $\mathcal{R}$,
given by
\begin{equation}
 P^\mathcal{R}(k) \equiv \frac{2\pi^2}{k^3} A_*^s \left(\frac{k}{k_*} \right)^{n_*^s -1},
\end{equation}
with $k_*$ the pivot scale with related primordial scalar amplitude $A_*^s$ and scalar spectral index $n_*^s$.
During matter domination, the relation 
\begin{equation}
\mathcal{R} = -\frac{5}{3}\Phi 
\end{equation}
is valid. Hence, the primordial power spectrum of $\Phi$ is given by
\begin{equation}
 P^\Phi (k) = \frac{9}{25}\frac{2\pi^2}{k^3} A_*^s \left(\frac{k}{k_*} \right)^{n_*^s -1}.
\end{equation}

Figure~\ref{recon_powers} shows the predicted data power spectrum using $R^{(2)}(r=\mathrm{const.})$ without instrumental beam, noise, or mask.
Having this response, we are able to construct the (data-space version of the) Wiener filter formula (see App.~\ref{app:wiener} for details),
\begin{equation}\label{wiener}
m^{(2)}(r)=P_\ell^\Phi (r) R^{(2) \dag}(r) \left[\tilde{C}^{TT}  + N \right]^{-1}d,
\end{equation}
with $P_\ell^\Phi (r)$ the primordial power spectrum projected onto the sphere at comoving distance $r$ and 
$\tilde{C}^{TT} = R P^\Phi R^\dag = M B C^{TT} B^\dag M^\dag$ where
\begin{equation}
\label{spec}
C^{XY}_\ell =\frac{2}{\pi} \int dk ~k^2 P^\Phi(k) {{g^X}_\ell(k)}{{g^Y}_\ell(k)}. 
\end{equation}
$X,Y$ can denote temperature $T$ or polarization $E-$mode.
Equation (\ref{wiener}) provides an optimal estimator of $\Phi_{\ell m}(r)$ and was stated first\footnote{For a detailed derivation see 
App. \ref{app:response_der} and \ref{app:uncertain}.} in
Ref.~\cite{2005PhRvD..71l3004Y}. The huge advantage of this method is the
reduction of computational time, by separating the whole inverse problem into many independent
distance-dependent ones. This method permits an easy parallelization of the Wiener
filter\footnote{The matrix inversion within Eq.~(\ref{wiener}), often solved by Krylov subspace methods
like the conjugate gradient method, is often computationally (very) expensive.} in the three-dimensional space.  
The $1\sigma$ uncertainty of this estimate, $\Delta m^{(2)}(r)$, is given by \cite{2009PhRvD..80j5005E}
\begin{equation}
\label{uncertainty}
\begin{split}
\Delta m^{(2)}(r)\equiv&~ \pm\sqrt{\mathrm{diag}\left[D\right]}\\
  =&~ \pm\sqrt{\mathrm{diag}\left[P^\Phi_\ell - P^\Phi_\ell R^{(2) \dag}\left(\tilde{C}^{TT} + N\right)^{-1} R^{(2)} P^\Phi_\ell \right]},
\end{split}
\end{equation}
where we have introduced the posterior covariance $D$ in data space. A proxy of this formula, used in our numerical calculations, can be found in App.~\ref{app:uncertain}.

\subsection{Temperature and polarization}
With future data releases of current experiments like Planck \cite{2014A&A...571A..16P}, it should be possible to include polarization data (P) with acceptable signal-to-noise level into considerations. 
Including polarization measurements, parametrized by the Stokes parameters $Q$,and $U$, the data are given by
\begin{equation}
 d= \begin{pmatrix}d^T\\d^Q\\d^U\end{pmatrix} = R\Phi + \begin{pmatrix}n^T\\n^Q\\n^U\end{pmatrix} 
\end{equation}
with corresponding response
\begin{equation}
 R = \underbrace{\begin{pmatrix}M_T B& 0&0\\0&M_P B&0\\0&0&M_P B\end{pmatrix}  W^{T,E}_{T,Q,U} }_{\equiv R^{T,E}_{T,Q,U}}
 \begin{pmatrix}R^T\\R^E\\0\end{pmatrix},
\end{equation}
where $R^{T,E}$ captures the radiation transfer, i.e.,
\begin{equation}
\left(R^{T,E}\Phi\right)_{\ell m} \equiv  \frac{2}{\pi}\int dk ~k^2 \int dr ~r^2 \Phi_{\ell m}(r) g^{T,E}_\ell(k)j_\ell(kr).
\end{equation}
The adiabatic radiation transfer functions are $g^{T,E}$ for temperature and E-mode polarization, respectively. For the formal definition of $g^{T,E}$ see, e.g.,
Refs.~\cite{1997PhRvD..55.1830Z,2001PhRvD..63f3002K}.
The operator $W^{T,E}_{T,Q,U}$ transforms a vector, containing temperature and E-mode polarization, into Stokes $I,Q,U$ parameters, which are directly
measured by experiments like WMAP or Planck. Therefore the generalized data-space version of the Wiener filter equation reads
\begin{equation}
\begin{split}
m^{(2)}(r) =&~  P_\ell^\Phi(r) \left( {R^{(2)}_T}^\dag(r)~{R^{(2)}_E}^\dag(r)~0  \right) 
\left(R^{T,E}_{T,Q,U}\right)^\dag \\
&~\times \left[R^{T,E}_{T,Q,U}
\begin{pmatrix}
C_\ell^{TT}&C_\ell^{TE}& 0\\
C_\ell^{TE}&C_\ell^{EE}& 0\\
0& 0& 0
\end{pmatrix}
\left(R^{T,E}_{T,Q,U}\right)^\dag + N \right]^{-1}\begin{pmatrix}d^T\\d^Q\\d^U\end{pmatrix},
\end{split}
\end{equation}
where $R^{(2)}_{X=T,E}$ denotes the two-dimensional version of $R^{X}$, analogous to Eq.~(\ref{R2}). The uncertainty is given analogously to Eq.~(\ref{uncertainty}).

The inclusion of polarization data will result in a significant improvement of reconstruction quality not least because $g^T_\ell (k)$ and $g^E_\ell (k)$ are 
out of phase and thus compensating the $\ell-$blind spots of each other,
which was also noticed by Ref.~\cite{2005PhRvD..71l3004Y} and can be observed in their Fig.~1. This is, however, only correct if the polarization data are not highly dominated by noise. 
\begin{figure}[ht]
\includegraphics[width=.5\textwidth]{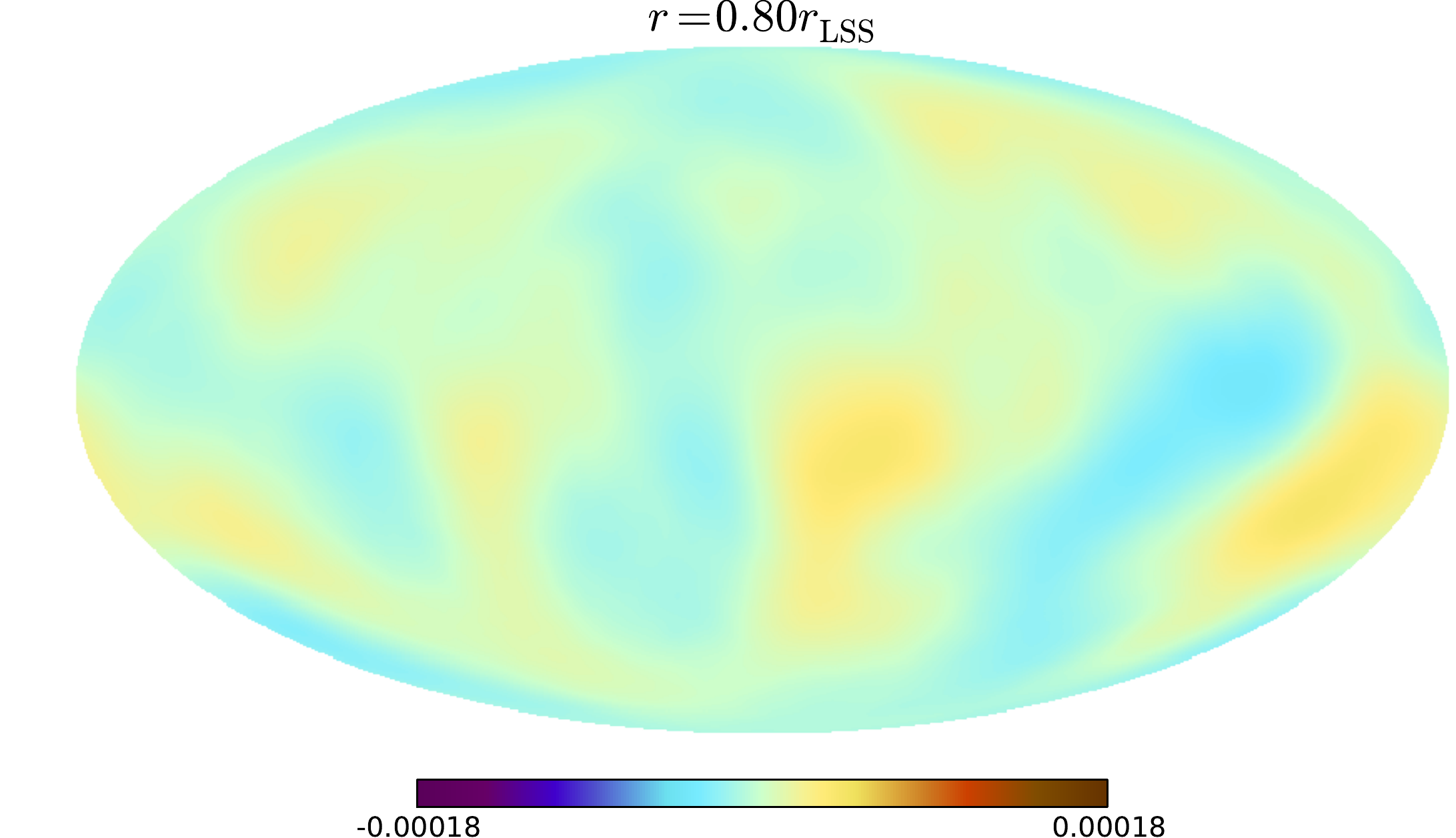}%
\includegraphics[width=.5\textwidth]{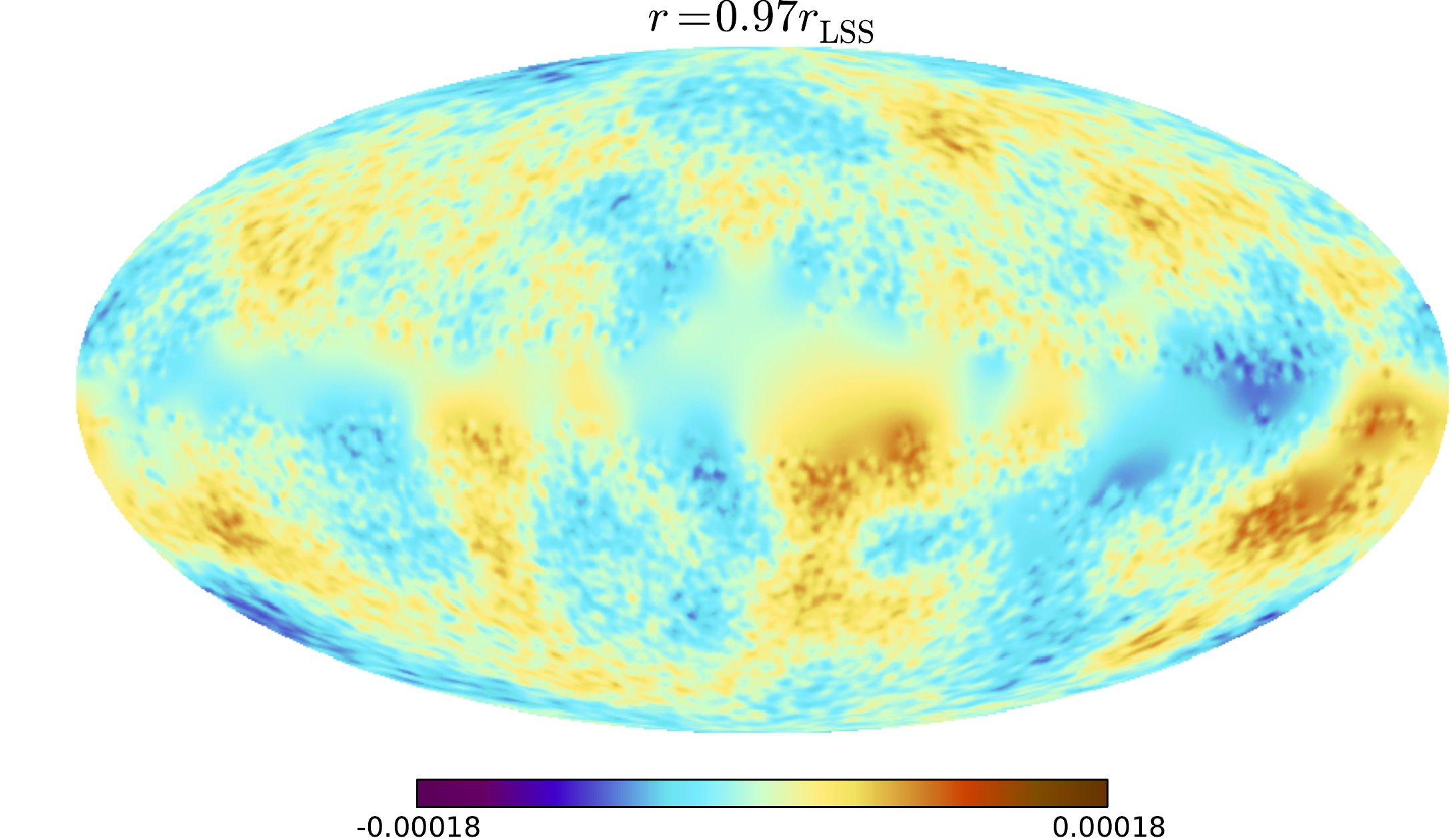}

\includegraphics[width=.5\textwidth]{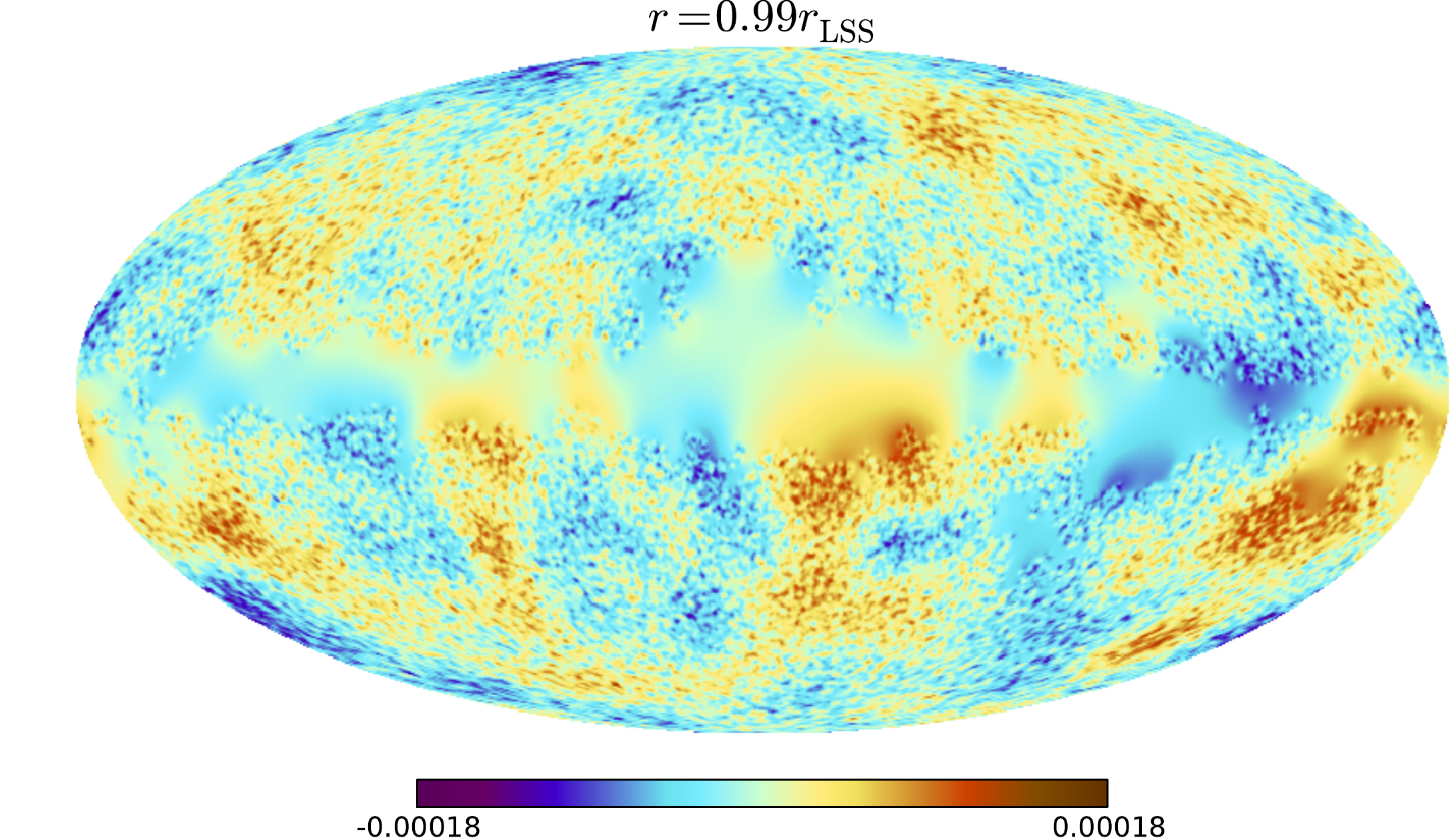}%
\includegraphics[width=.5\textwidth]{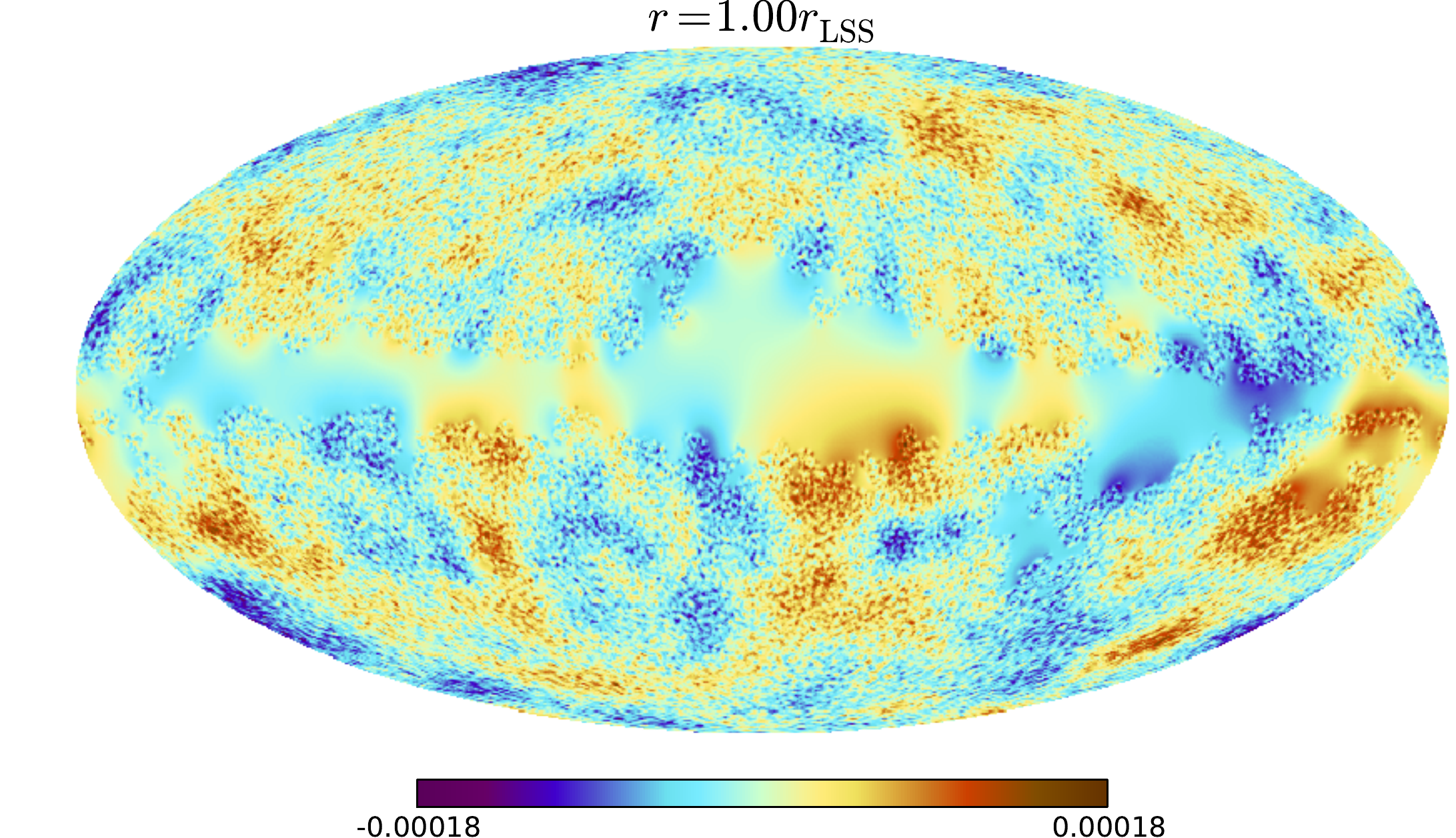}

\includegraphics[width=.5\textwidth]{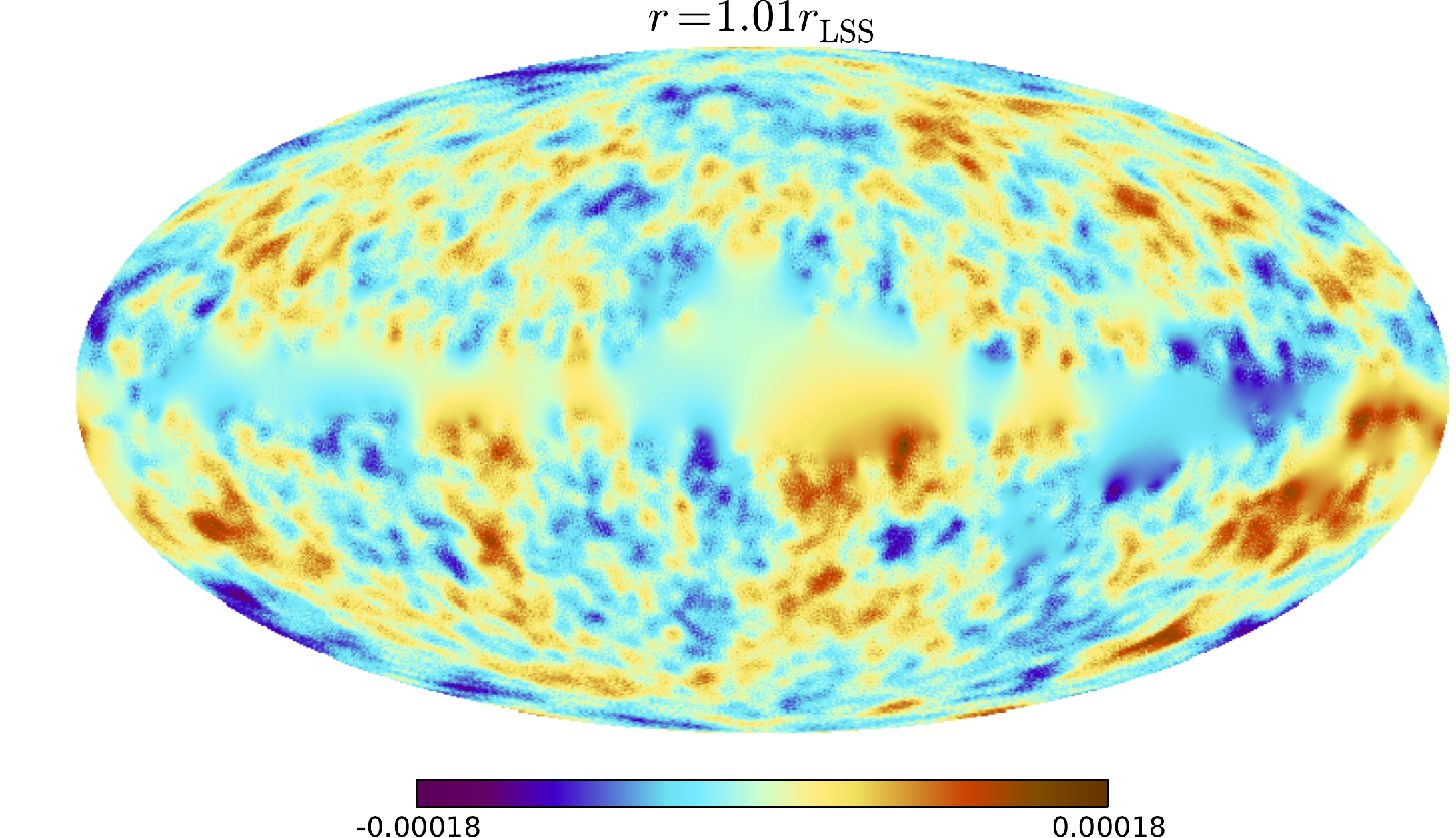}%
\includegraphics[width=.5\textwidth]{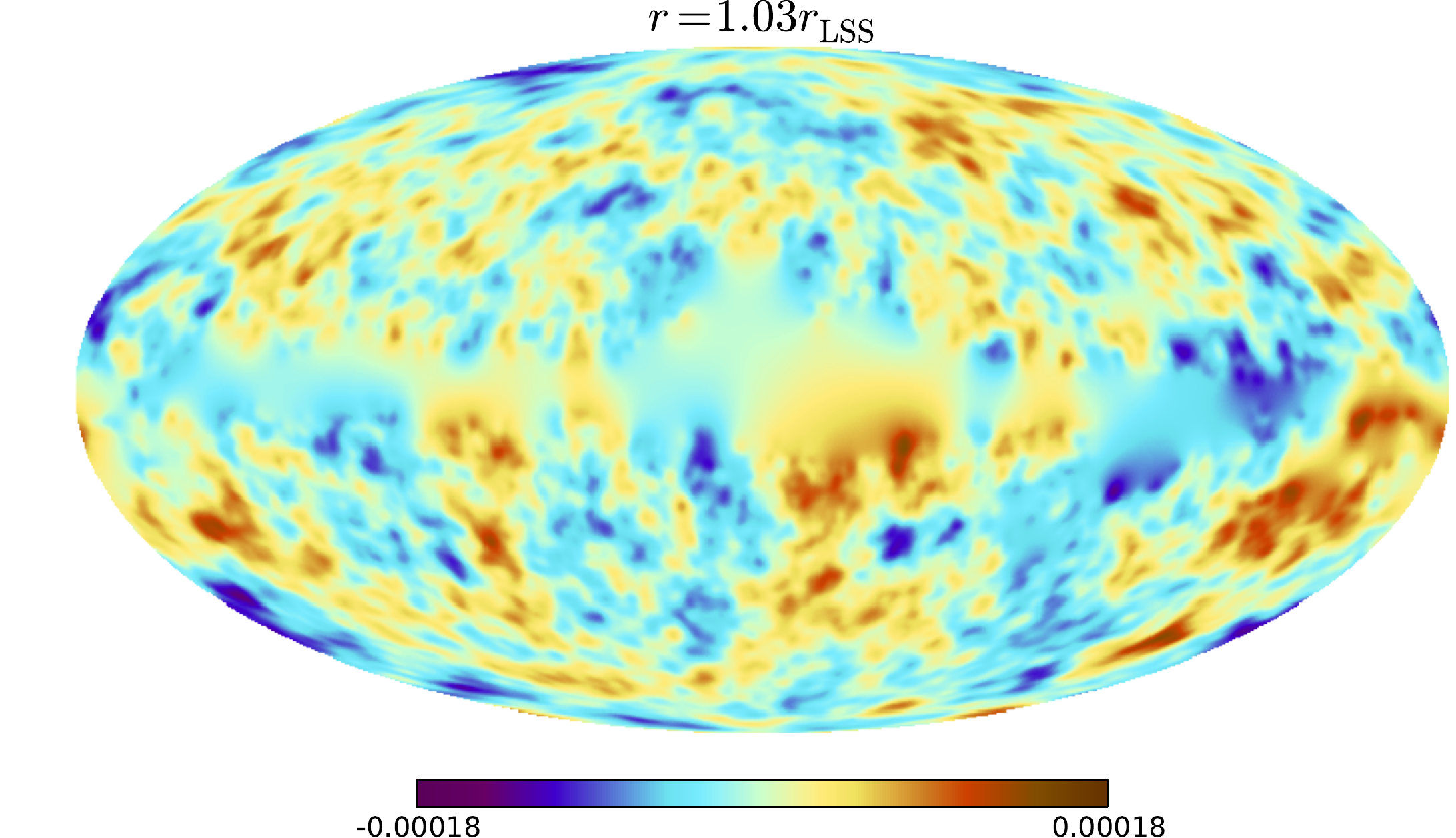}

\includegraphics[width=.5\textwidth]{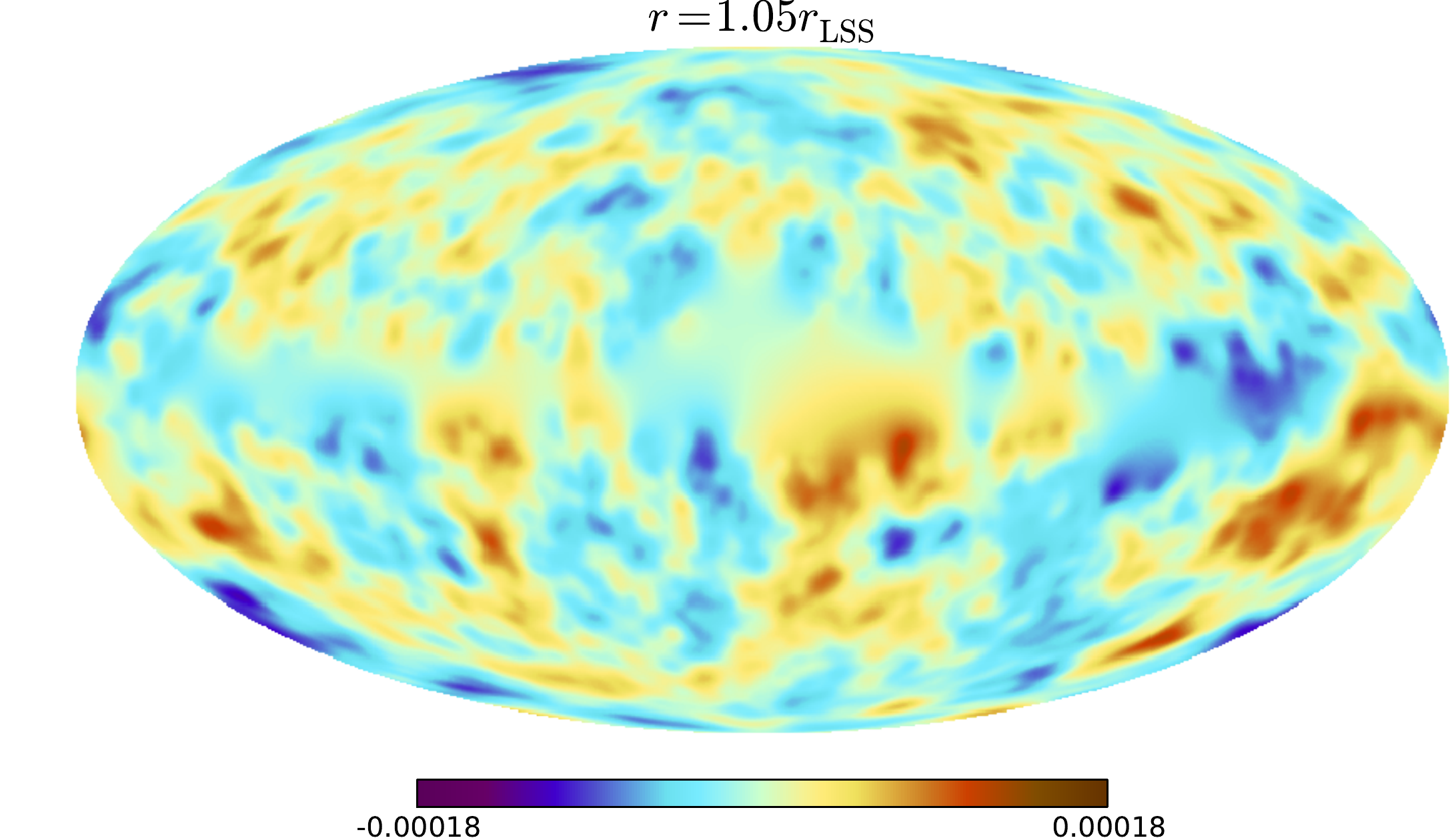}%
\includegraphics[width=.5\textwidth]{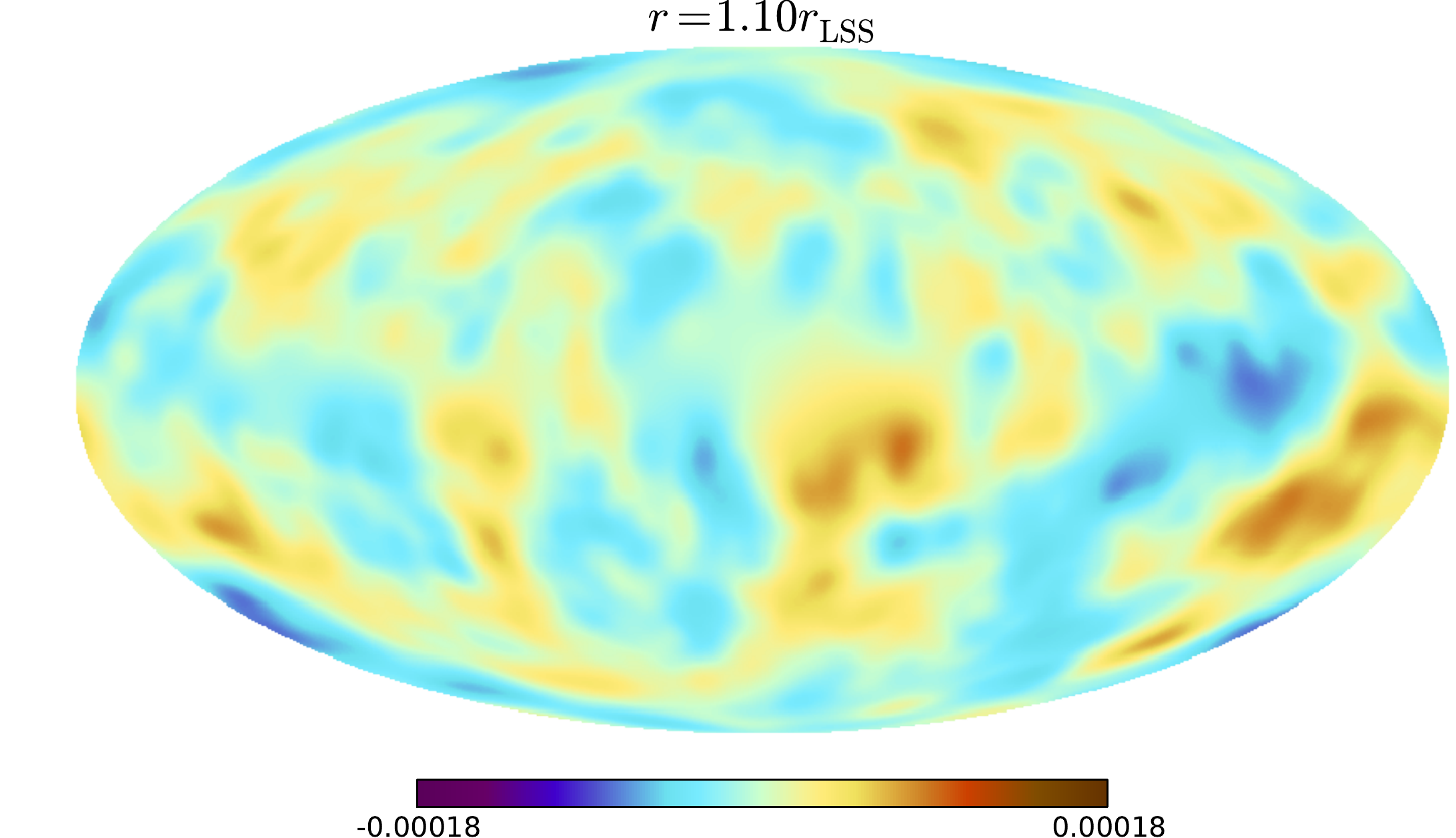}
 \caption{(color online) All-sky maps of the reconstructed primordial scalar potential at different
 comoving distances according to Eq.~(\ref{wiener}) in the vicinity of the recombination sphere with $r=r_\mathrm{LSS}$. A Mollweide projection is used.}%
\label{reconstr}
\end{figure}

\begin{figure}[ht]
\includegraphics[width=.5\textwidth]{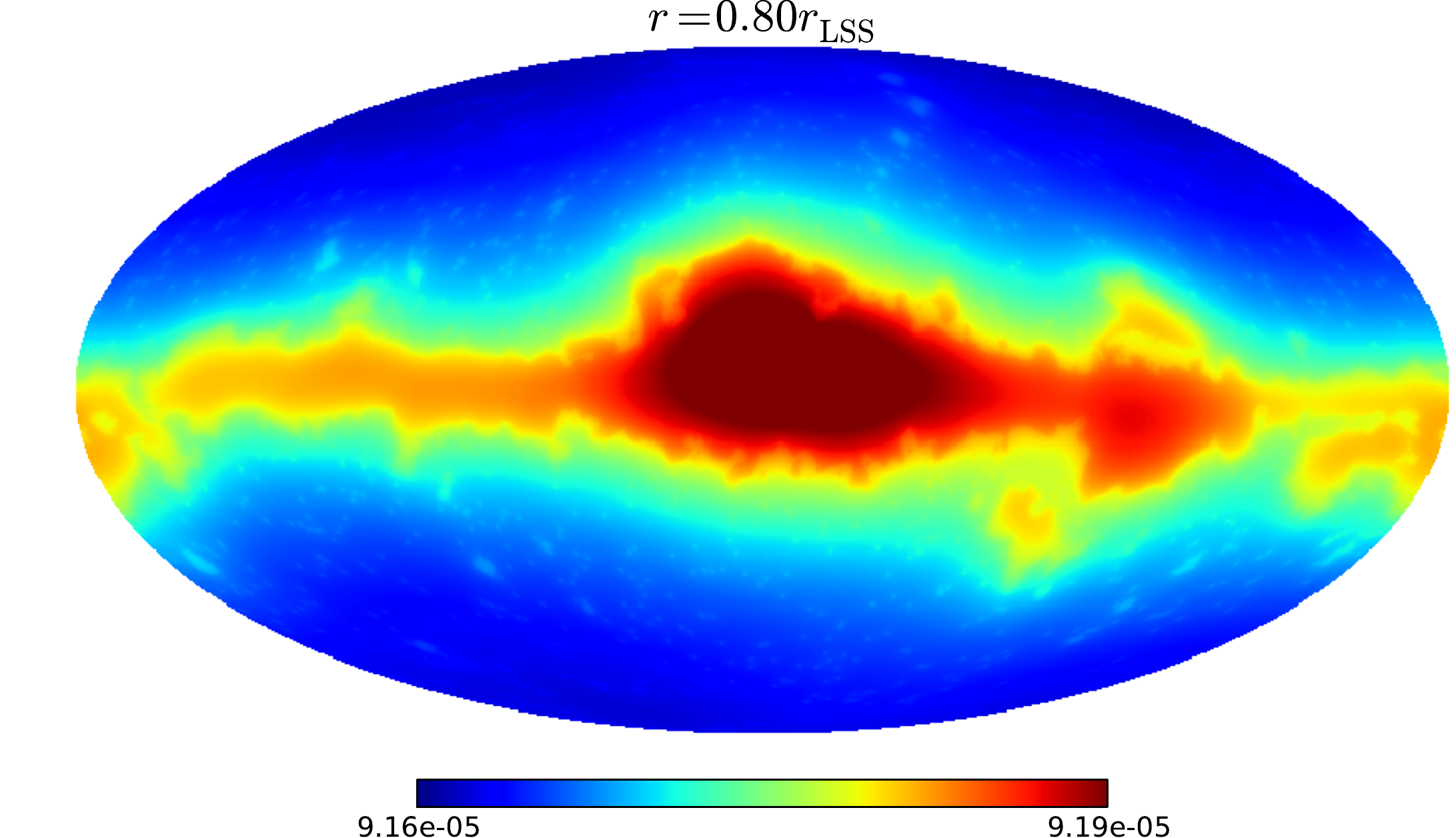}%
\includegraphics[width=.5\textwidth]{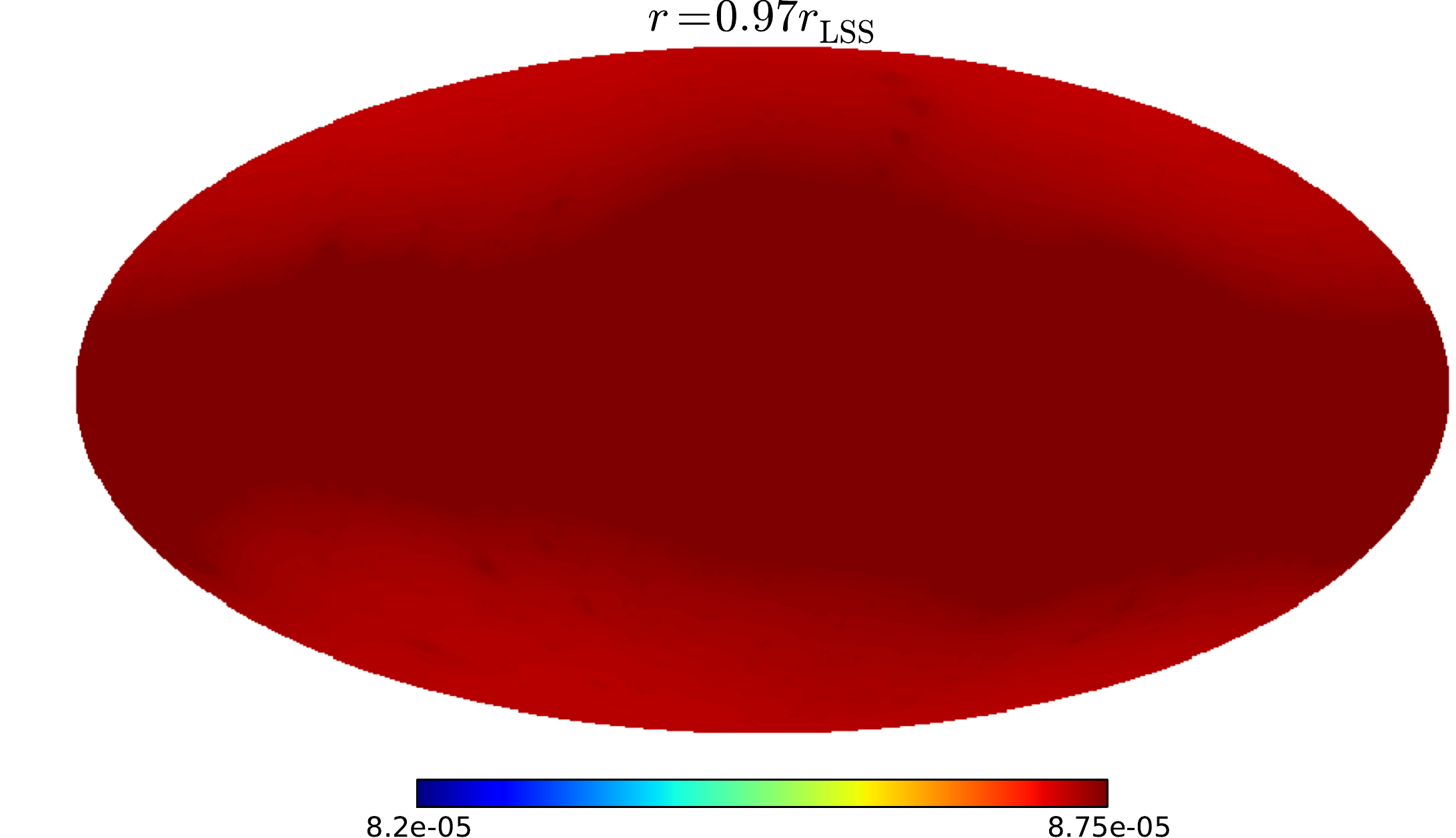}

\includegraphics[width=.5\textwidth]{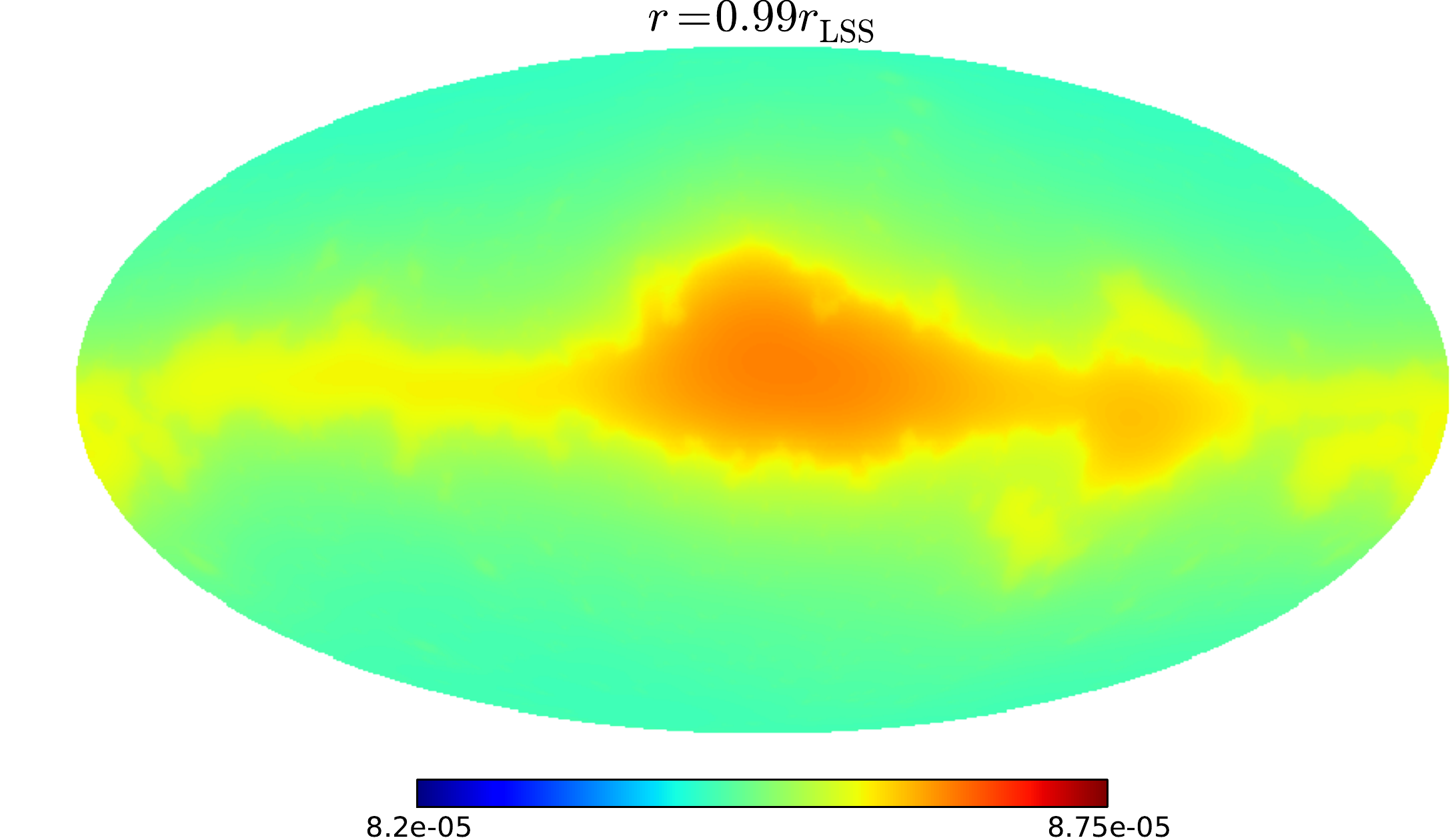}%
\includegraphics[width=.5\textwidth]{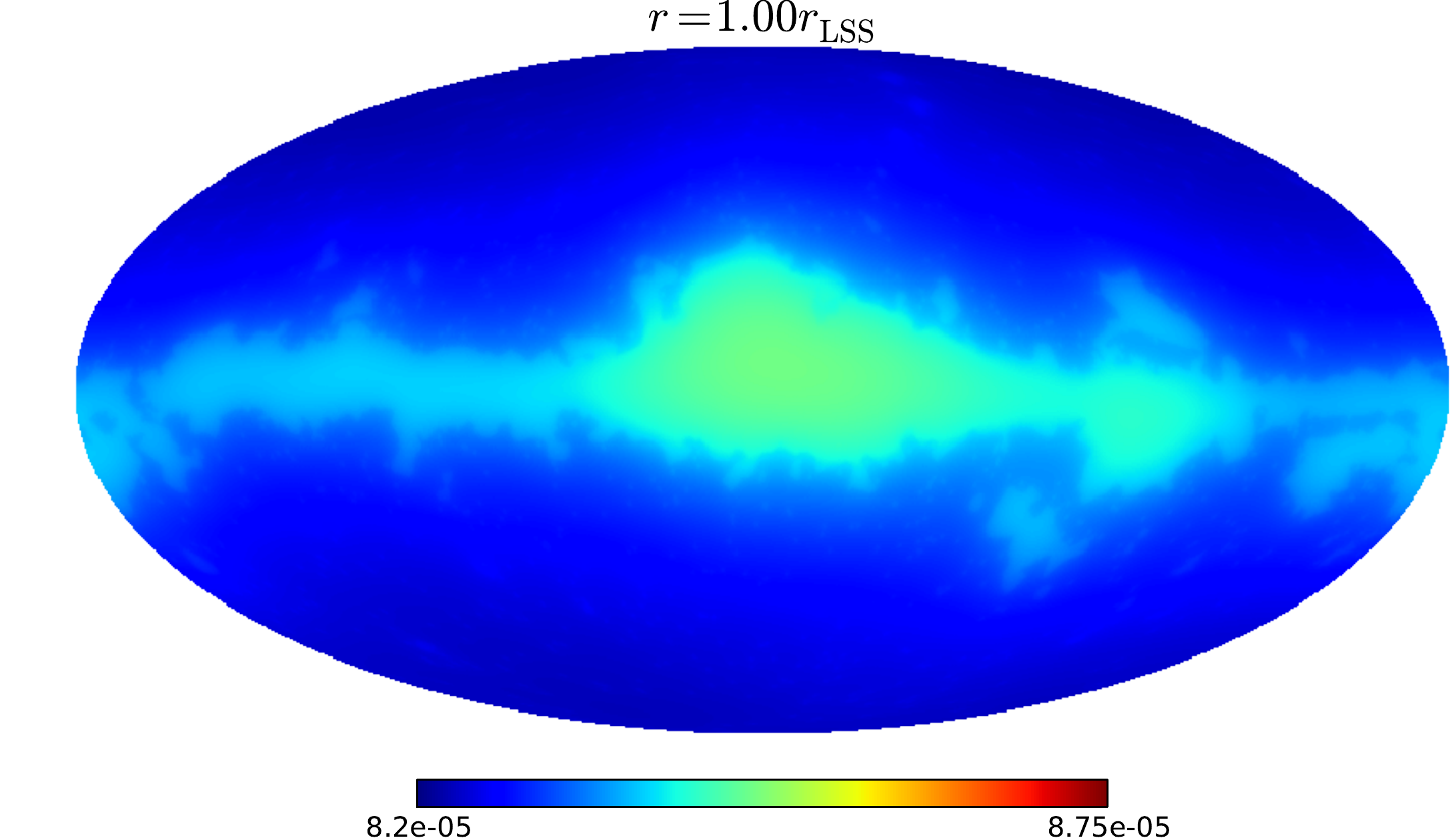}

\includegraphics[width=.5\textwidth]{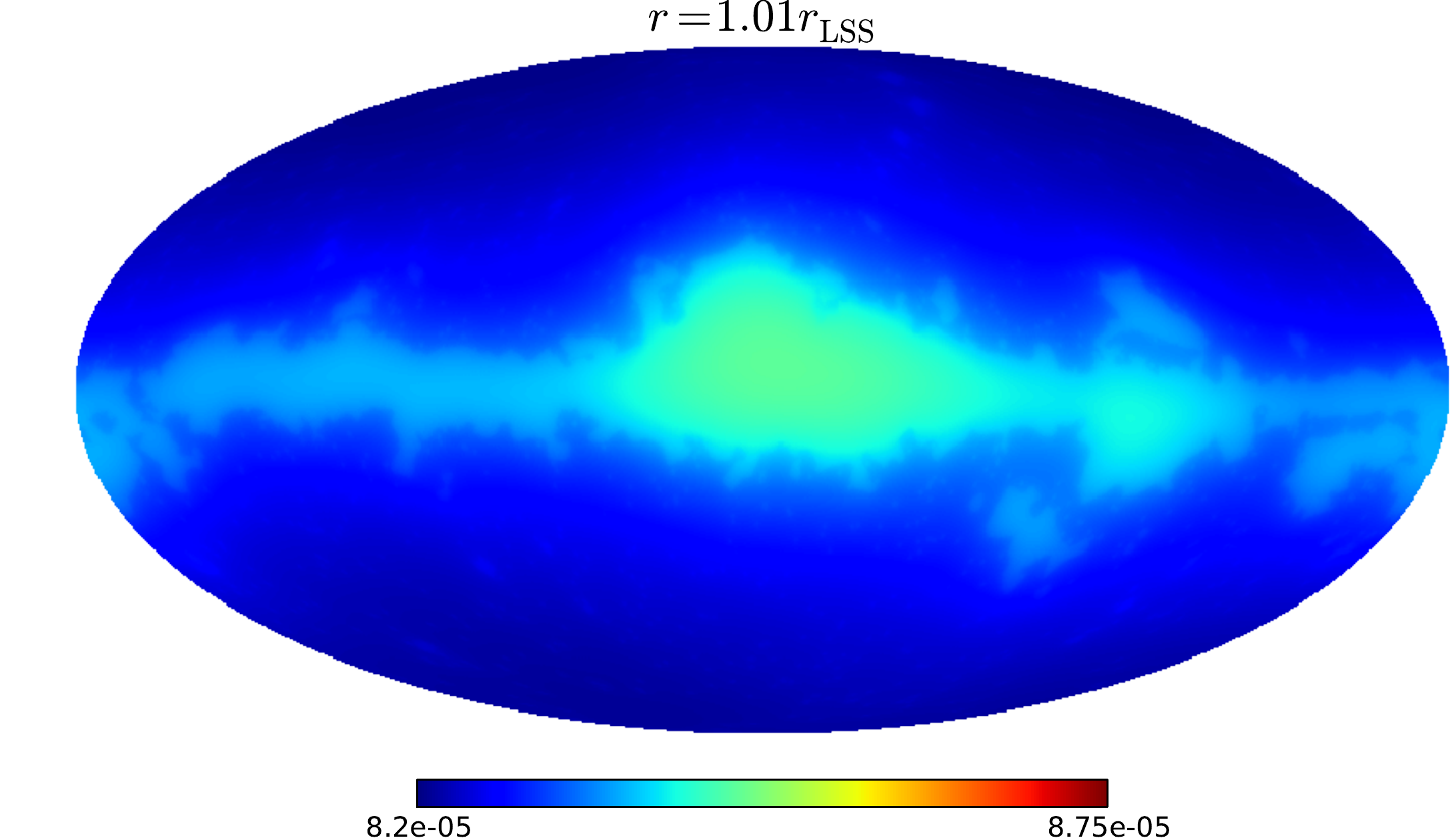}%
\includegraphics[width=.5\textwidth]{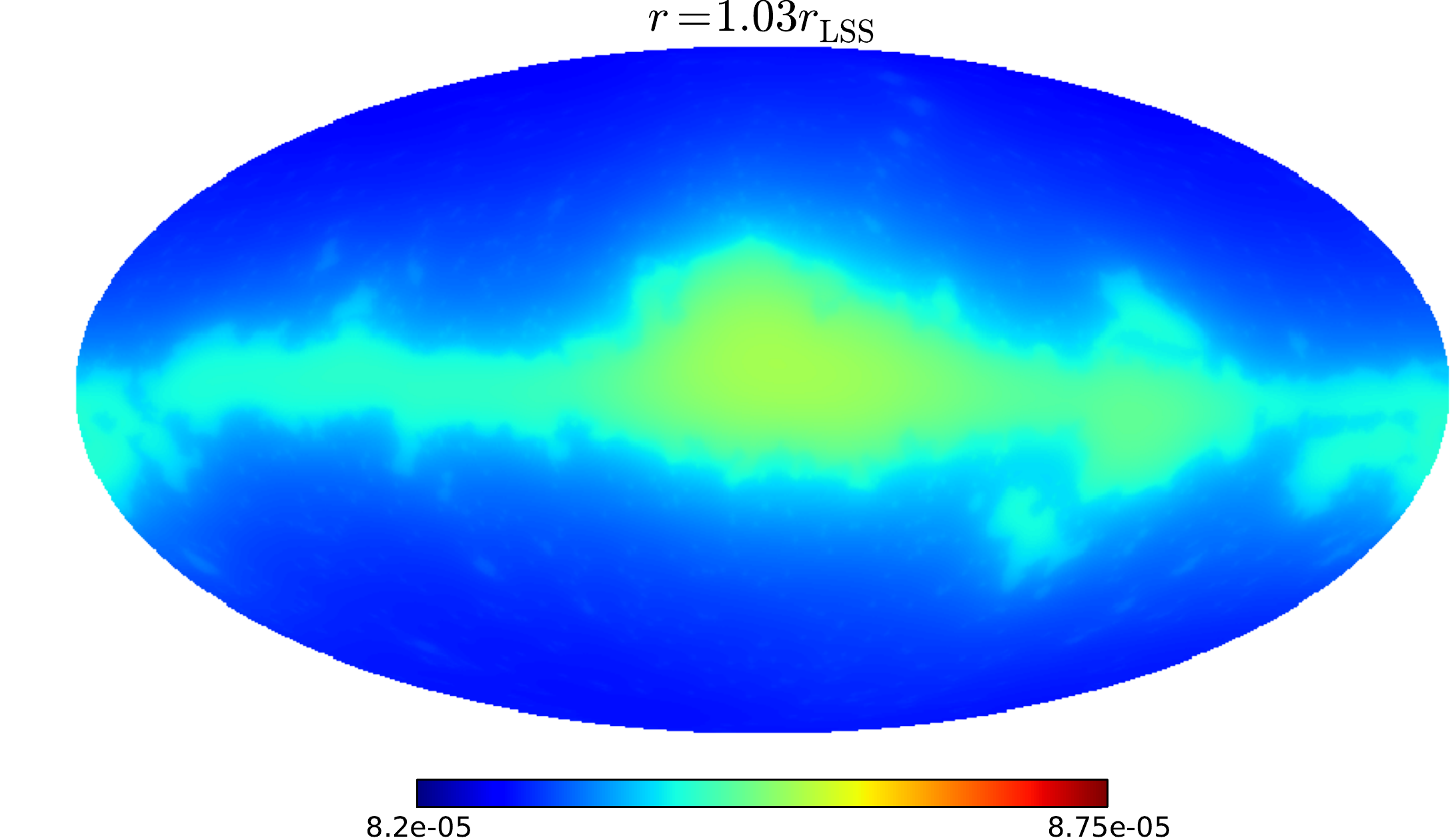}

\includegraphics[width=.5\textwidth]{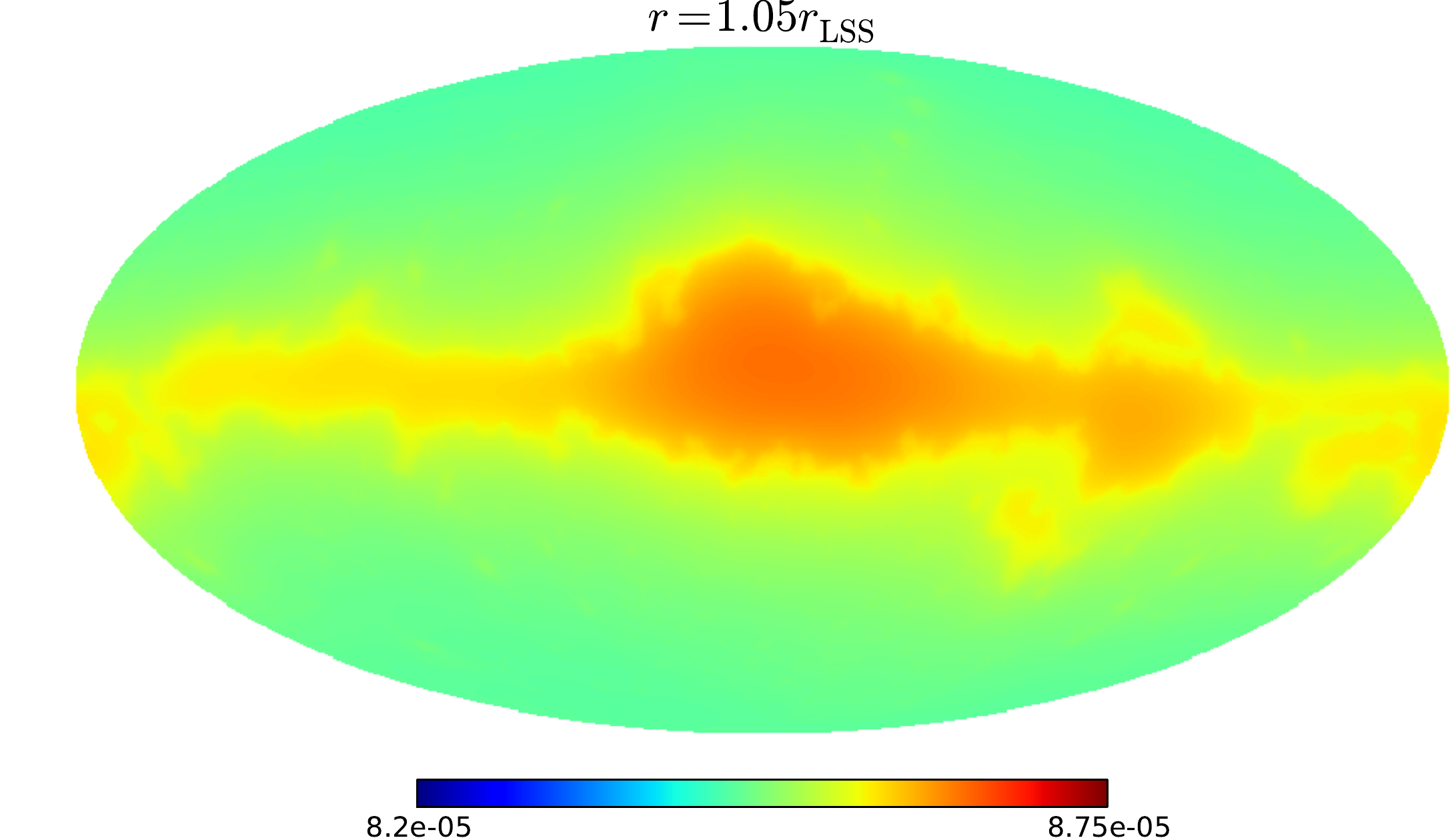}%
\includegraphics[width=.5\textwidth]{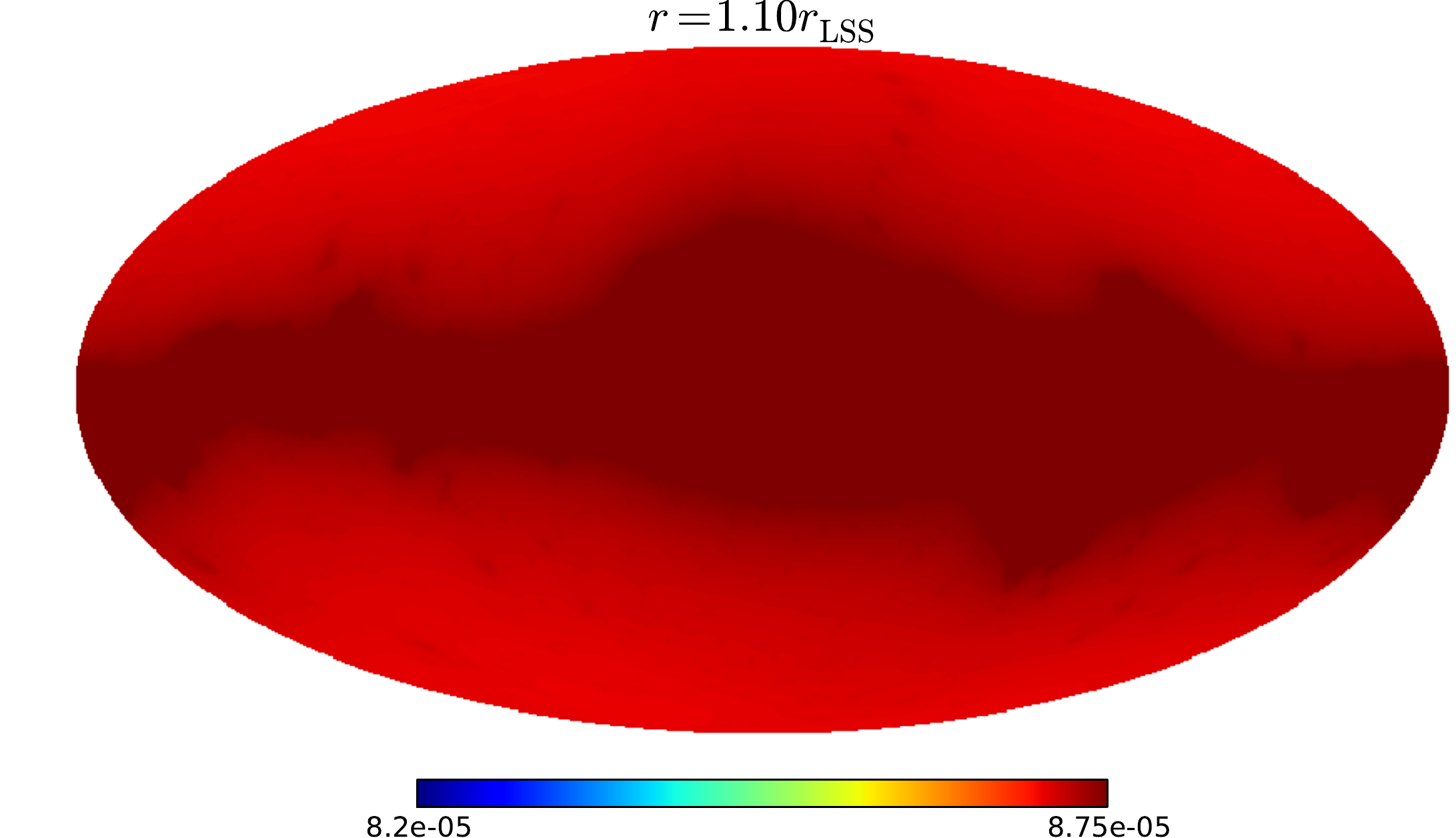}
 \caption{(color online) $1\sigma$ uncertainty maps of the corresponding all-sky maps of of Fig.~\ref{reconstr} according to Eq.~(\ref{sigma}) in the vicinity of the
 recombination sphere with $r=r_\mathrm{LSS}$. A Mollweide projection is used. Note that the color bar for $r=0.80r_\mathrm{LSS}$ is a different one,
 showing the natrual bounds of the uncertainty map. All uncertainty maps share this morphology.} %
\label{U_maps}
\end{figure}

\subsection{Primordial power spectrum reconstruction}\label{sec:crit}
Once a signal estimate (optimally with uncertainty) is available the power spectrum of the stochastic process underlying the signal
generation might be inferred. Usually, however, an initial guess of the signal power spectrum
is required to obtain a Wiener filter signal in the first place.
This initial guess spectrum can affect the spectrum estimate and therefore might act as a hidden prior. 
In order to forget the initial guess,
the procedure of signal and spectrum inference should be iterated until it has converged onto a spectrum
that is then independent of the initial starting value.
Fortunately,
the primordial power spectrum is constrained well by the existing\footnote{See section 7 of Ref.~\cite{2014A&A...571A..22P} and Ref.~\cite{2014A&A...566A..77P} for
an overview of the literature on such methods.} CMB data-sets so that this process should converge rapidly.
This iterative, unparametrized method was derived in Refs.~\cite{2011PhRvD..83j5014E,2012arXiv1210.6866O} and named critical filter.
It can be regarded as a maximum a posteriori estimate of the logarithmic power spectrum and the
assumption of a scale invariant Jeffreys prior of its amplitudes.
The power spectrum on the sphere is written as
\begin{equation}
P^{\Phi}_{\vec{\ell}{\vec{\ell}}'} = \delta_{{\vec{\ell}} {\vec{\ell}}' } P_\ell^\Phi~~\mathrm{with}~~\vec{\ell}\equiv (\ell,m).
\end{equation} 
The iterative critical filter formula including a spectral smoothness prior is then given by Eq.~(\ref{wiener}) and
\begin{equation}
\label{crit_smooth}
P_\ell^\Phi = \frac{\sum_{\{\vec{\ell}'|\ell' = \ell \}}\left(m_{\vec{\ell}}^{(2)}  m_{\vec{\ell}'}^{(2) \dag} + D_{\vec{\ell} \vec{\ell}'} \right)}{\rho_\ell + 2(S\ln P^\Phi)_\ell}, 
\end{equation}
where $\rho_\ell = \sum_{\{\vec{\ell}'|\ell' = \ell \}} 1$ is the number of degrees of freedom on the multipole $\ell$ and $S$ an operator that
enforces smoothness (for details see Ref.~\cite{2012arXiv1210.6866O}).

\begin{figure}[ht]
\includegraphics[width=.5\textwidth]{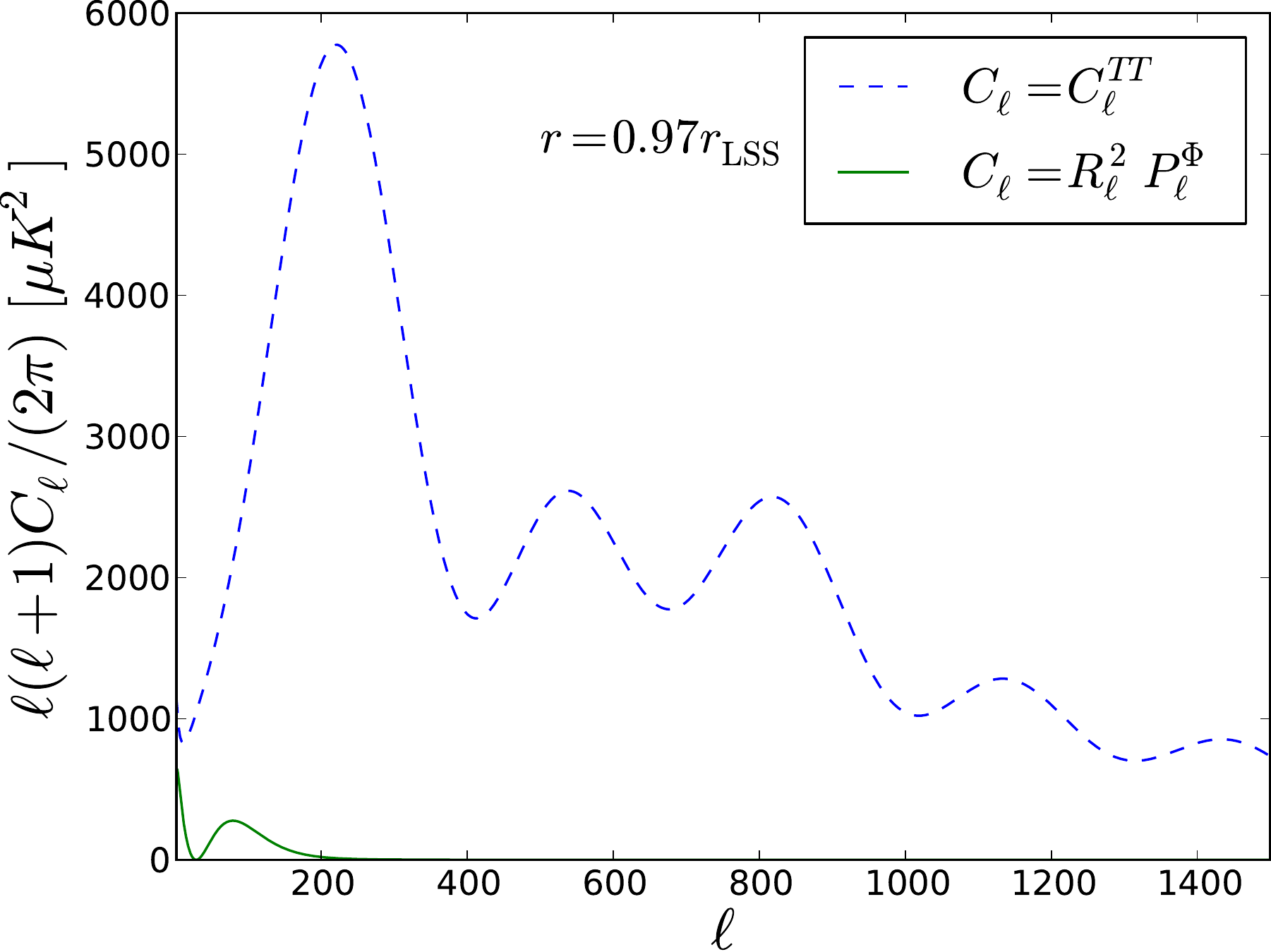}%
\includegraphics[width=.5\textwidth]{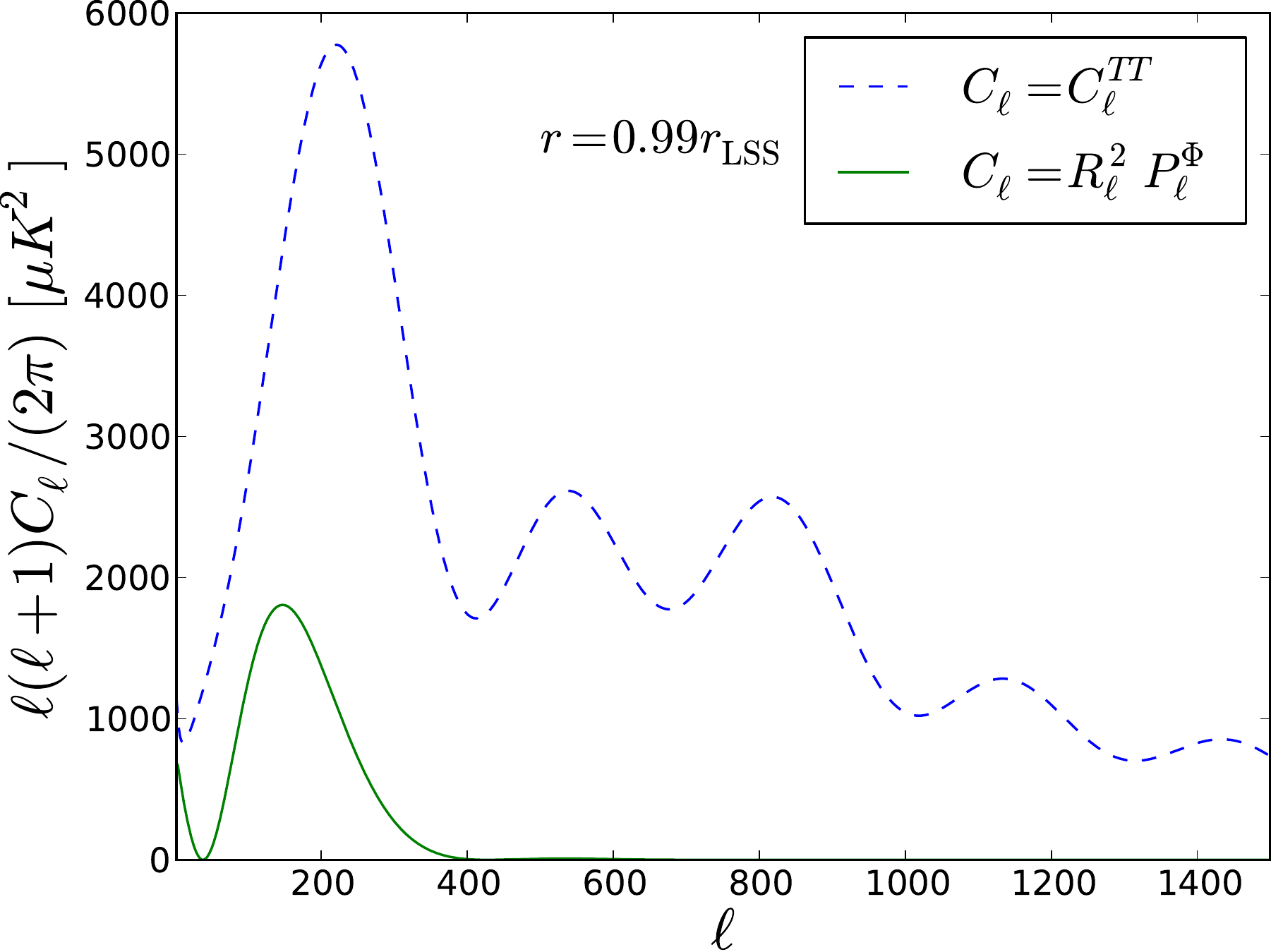}

\includegraphics[width=.5\textwidth]{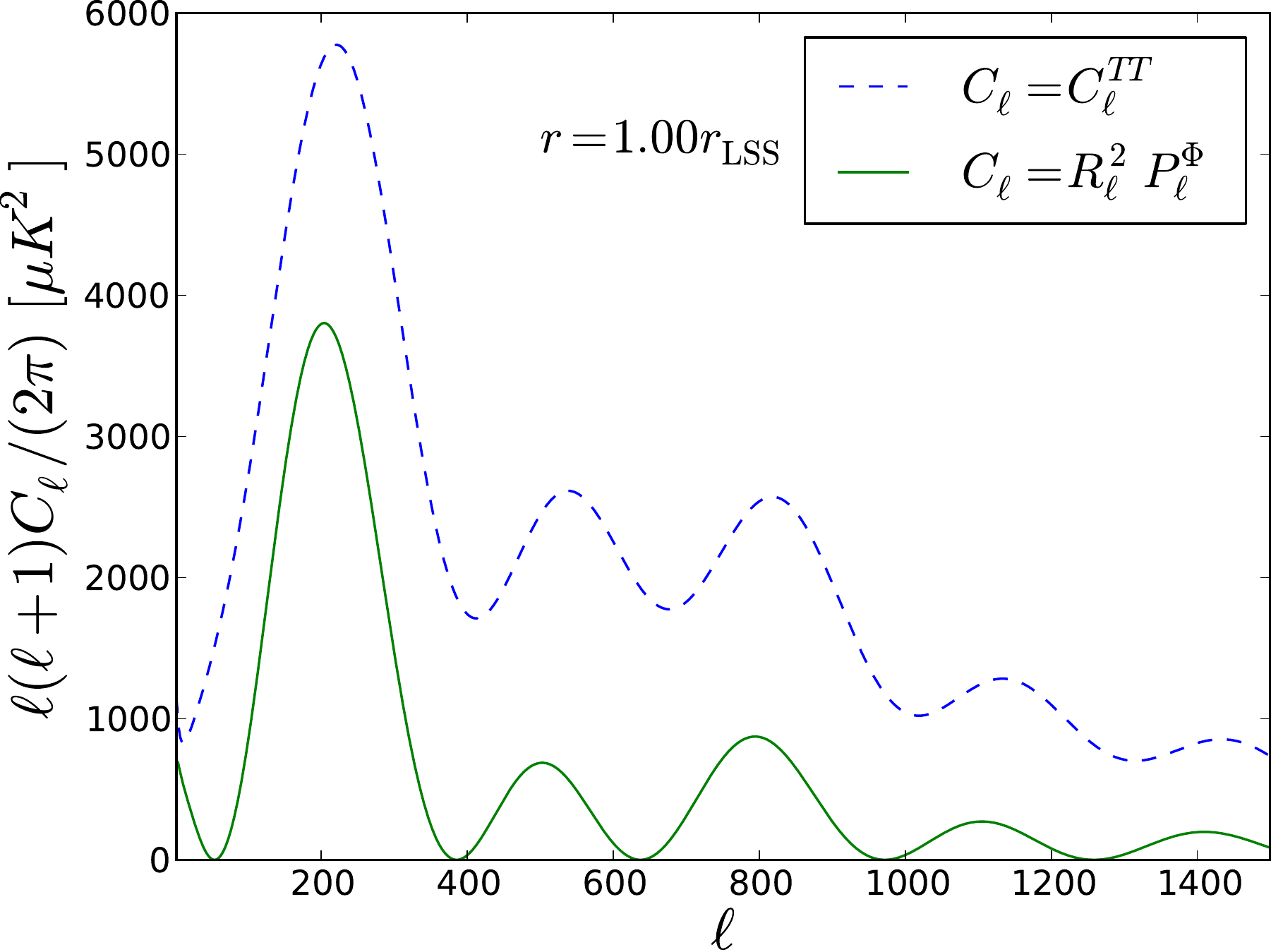}%
\includegraphics[width=.5\textwidth]{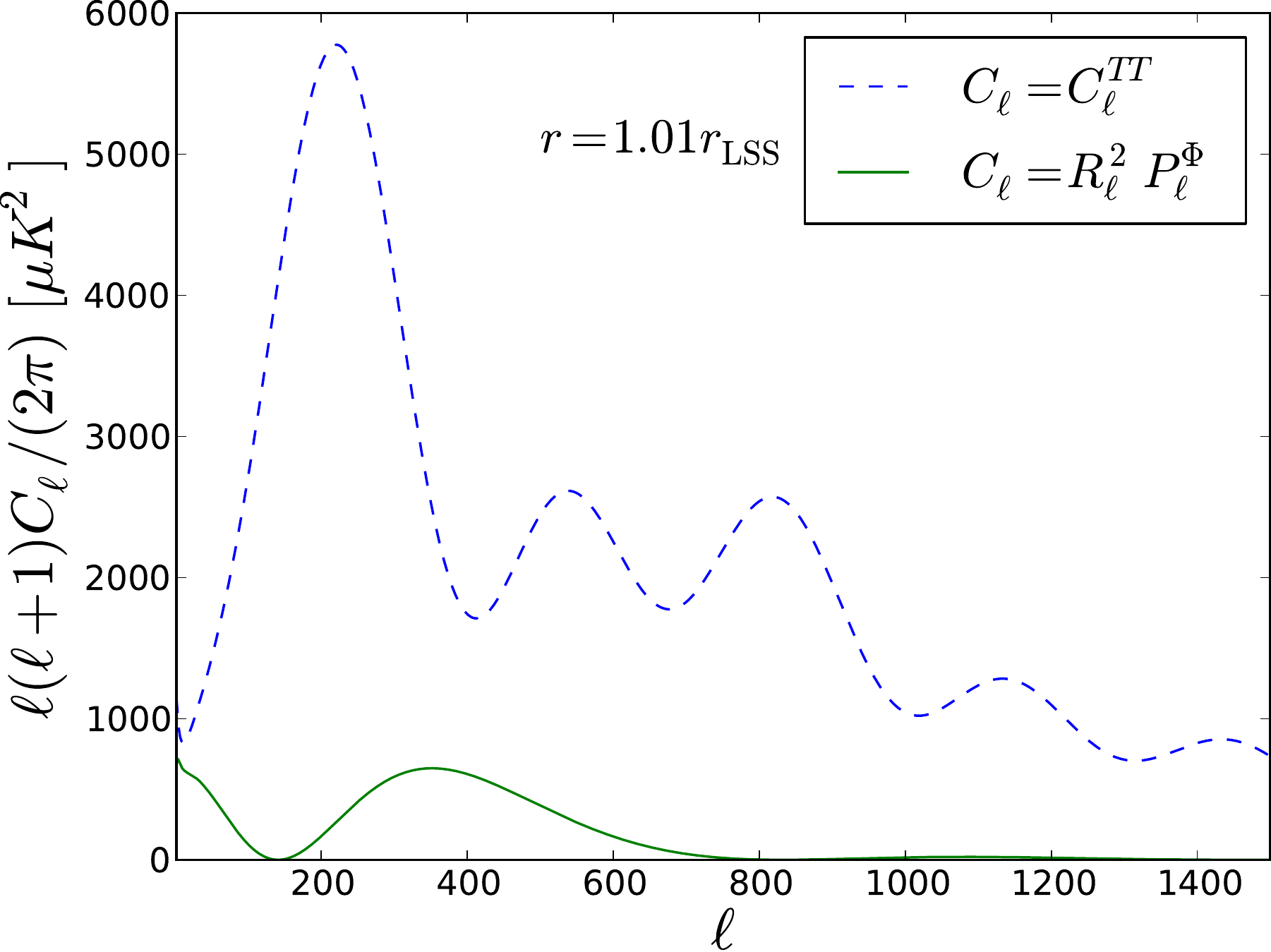}

\includegraphics[width=.5\textwidth]{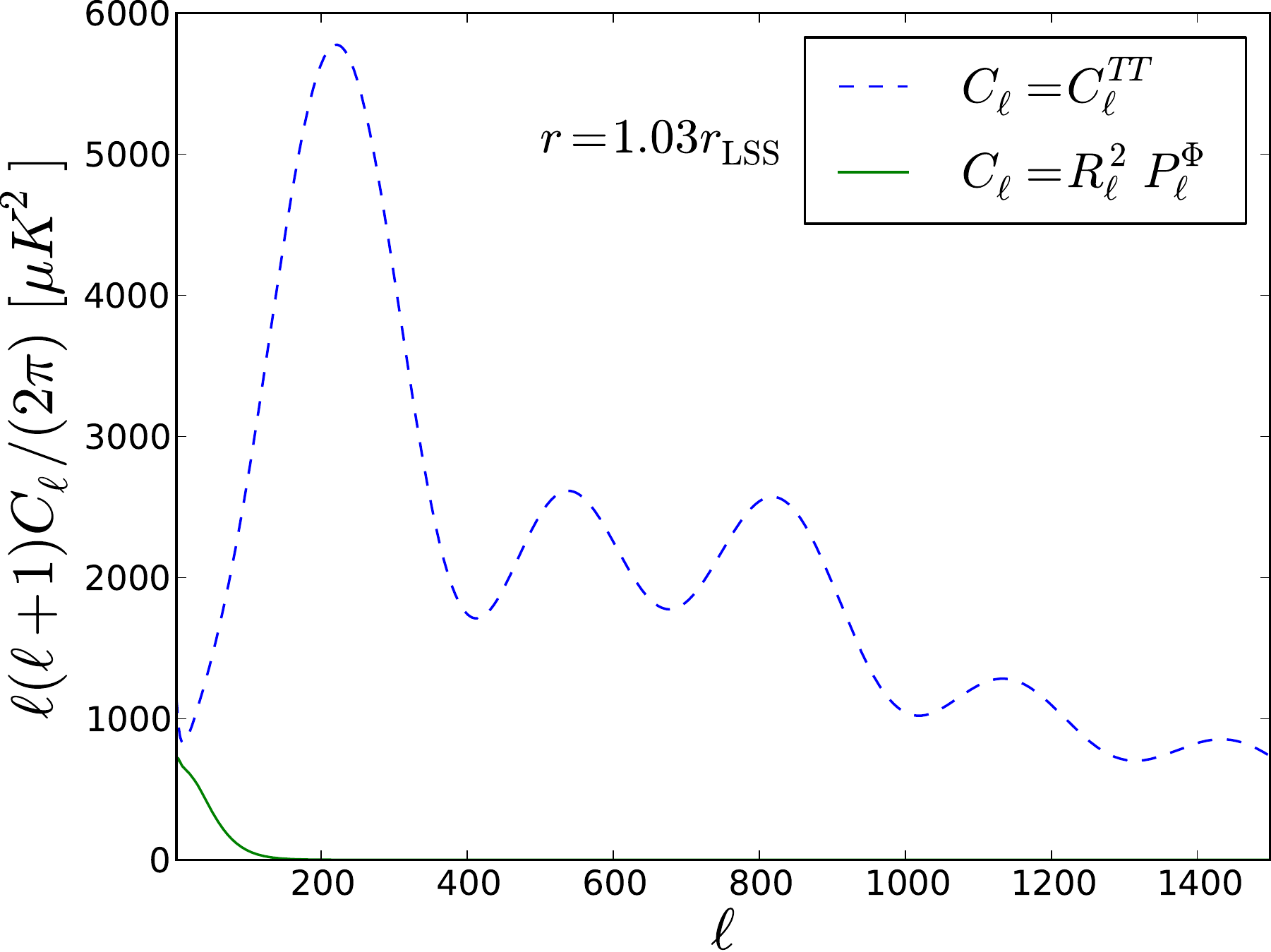}%
\includegraphics[width=.5\textwidth]{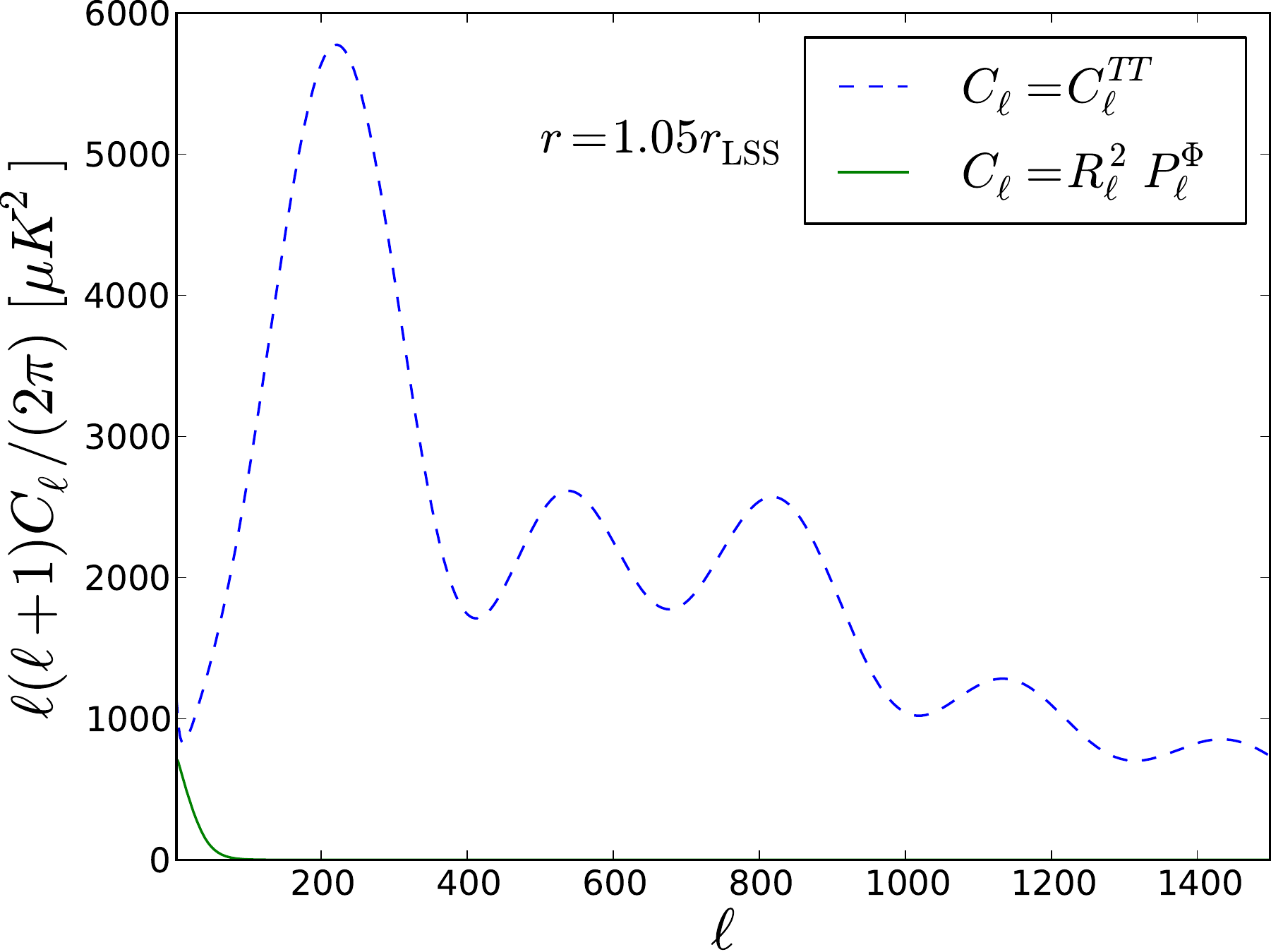}
 \caption{(color online) Predicted power spectra of data simulated with the estimator response $R^{(2)}$ compared to the CMB data power spectrum. 
 The $\ell-$blind spots move from large scales at distances $r<r_\mathrm{LSS}$ to small scales at $r>r_\mathrm{LSS}$. The amplitude of the predicted power spectra gets maximal at $r=r_\mathrm{LSS}$. 
 For clarity and comprehensibility we exclude the instrumental beam, noise, and observational mask.}%
\label{recon_powers}
\end{figure}
\section{Temperature-only reconstruction of the primordial scalar potential}\label{sec:recon}
\subsection{Input values and settings}
We analyze the full resolution ($\mathrm{nside = 512}$) coadded nine-year WMAP (foreground-cleaned) V-band frequency temperature map, masked with the primary
temperature analysis mask (KQ85: 74.8\% of the sky). The data as well as the corresponding beam transfer function and noise properties (see App.~\ref{app:noise})
we used can be found at \url{http://lambda.gsfc.nasa.gov/product/map/dr5/m_products.cfm} \cite{2013ApJS..208...20B,2013ApJS..208...19H}.
We did not take polarization data into considerations due to the suboptimal signal-to-noise levels. To be consistent with the WMAP team's measurements
we use the cosmological parameters obtained by their data analysis to compute the radiation transfer function as well as the primordial power spectrum. In particular this has been done by using
\texttt{gTfast}\footnote{\url{http://www.mpa-garching.mpg.de/~komatsu/CRL/nongaussianity/radiationtransferfunction/}}, which is based
on \texttt{CMBFAST}\footnote{\url{http://lambda.gsfc.nasa.gov/toolbox/tb_cmbfast_ov.cfm}} \cite{1996ApJ...469..437S}. We used the following settings:
pivot scale $k_*=0.002~\mathrm{Mpc}^{-1}$, spectral index $n_*^s=0.962$,
spectral amplitude $A_*^s = 2.46\times 10^{-9}$, noise level $\sigma^{\mathrm{V-band}}_0 = 3.131\times 10^{-3}$ K, CMB temperature $T_\mathrm{CMB} = 2.726$ K,
optical depth $\tau = 0.088$, density parameters $\Omega_b = 0.046,~\Omega_c = 0.0231,~\Omega_\Lambda = 0.723$, Hubble constant $H_0 = 70.2 ~\mathrm{km/s/Mpc}$,
helium abundance $Y_\mathrm{He} = 0.24$, and the effective number of massless neutrino species $N_\nu^\mathrm{eff} = 3.04$. 
The resulting distance to the LSS amounts $1.40147\times 10^4$ Mpc.

\begin{figure}[ht]
\begin{center}
\includegraphics[width=.5\textwidth]{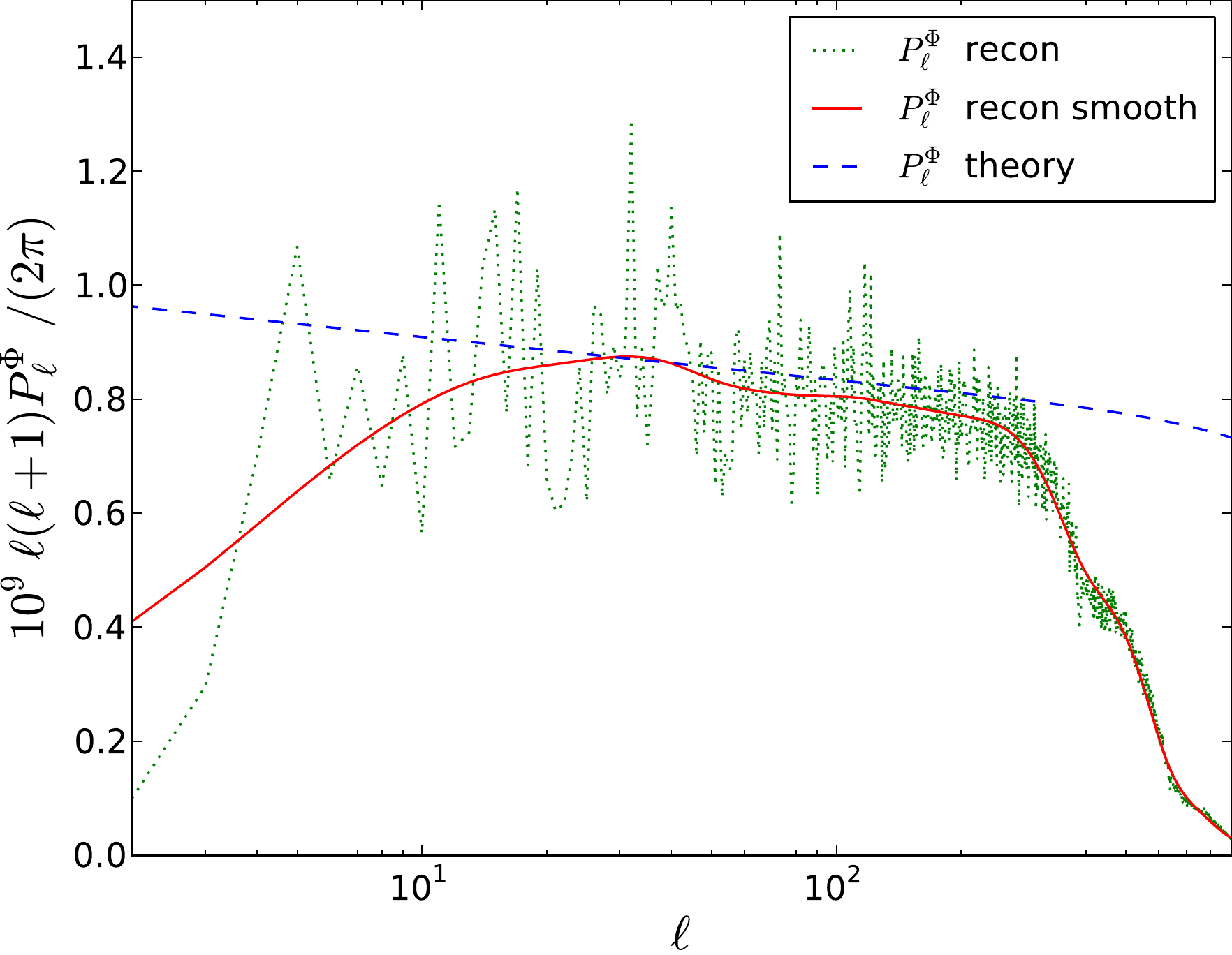}%
\includegraphics[width=.5\textwidth, height = 6.1cm]{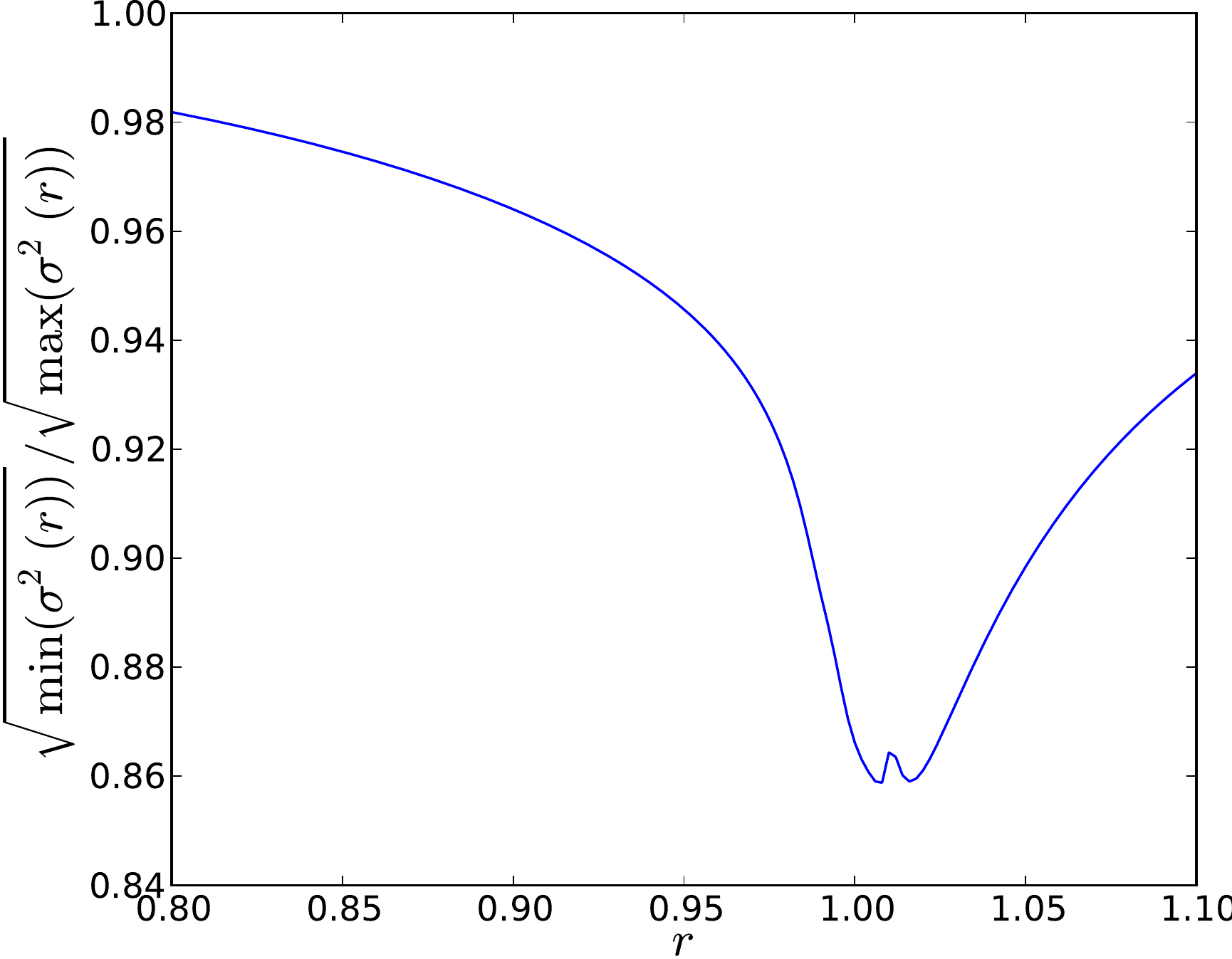}
\caption{(color online) Left: Estimated primordial power spectrum of $\Phi(r=r_\mathrm{LSS})$ according to Eq.~(\ref{crit_smooth}). Masking effects as well as the estimators power loss are compensated.
At scales smaller than $\ell \approx 300$ the reconstruction fails due to sub-horizon physics \cite{2005PhRvD..71l3004Y} and noise-dominance. Right: Relative $1\sigma$-uncertainty along the radial coordinate.
Minimal values of $\sigma$ correspond to Eq.~(\ref{sigma2}) with ``no mask'', maximal values to the same equation with ``all mask''.}%
\label{power_est}
\end{center}
\end{figure}

\subsection{Results}
With the parameters defined in the previous paragraph, we have reconstructed a shell around the last scattering surface ($0.8\times r_\mathrm{LSS}$ to $1.1\times r_\mathrm{LSS}$) in $151$ slices as well
as additional 6 slices within the range $(50\%-80\%)\times r_\mathrm{LSS}$ from real data, see Fig.~\ref{reconstr}. For all reconstructions $1\sigma$-uncertainty maps are provided, see Fig.~\ref{U_maps} 
as well as the relative $1\sigma$-error along the radial coordinate, see Fig.~\ref{power_est} (Right). 
A detailed description of the calculation of these uncertainty maps can be found in App.~\ref{app:uncertain}.
The respective data files of the reconstruction can be found at \url{http://www.mpa-garching.mpg.de/ift/primordial/}.
For the most interesting sphere
at $r=r_\mathrm{LSS}$ we also provide a power spectrum estimate, see Fig.~\ref{power_est} (Left). This power spectrum estimate has been obtained with
the critical filter formula with smoothness prior but without iterations\footnote{With the correct application
of the critical filter (iterative) one might
be able to detect features in the primordial power spectrum \cite{2014JCAP...06..048D}. 
This, however, would require a highly resolved data set including polarization to compensate for the $\ell-$bind spots (one cannot get rid of with temperature data only) with a high signal-to-noise 
level in $T$-, $Q$-, and $U$-data maps. Perfect candidates for such data sets are future CMB experiments
and Planck polarization data releases.} and $D$ set to zero (defined in Eq.~(\ref{uncertainty})). 

We also phenomenologically\footnote{The power-loss is corrected by convolving the reconstructed $\Phi$ with
$\alpha_\ell \equiv \sqrt{C_\ell^{TT}/ (R_\ell^2 P^\Phi_\ell) }~~\forall \ell: R^2_\ell P^\Phi_\ell \neq 0$
before performing the power spectrum estimation. We also investigated how the mask affects the power spectrum of $\Phi$,
by calculating $\beta_l\equiv \left\langle \mathrm{power}\left[R^\mathrm{mask}(\Phi)\right]/P^\Phi_\ell\right\rangle$ where $\mathrm{power}[~.~]$ denotes the application of
the critical filter formula with smoothness prior. We re-scaled the inferred power spectrum with $1/\beta_\ell$.} corrected for the effect of masking and power-loss in the predicted power spectra of data
simulated with the estimator response $R^{(2)}$ in comparison to the power spectrum of Eq.~(\ref{spec}). 
Therefore our spectrum estimate should rather be regarded as providing a consistency check of the algorithm
than to necessarily provide precisely the cosmological power spectrum. 
Having stated these caveats, we like to note that a deviation from the power-law primordial power spectrum
is not apparent over roughly one order of magnitude in Fourier space.  

Some of the reconstructed slices of the primordial scalar potential might look suspiciously crumby at first.
The reason for this property are the $\ell-$blind spots in the response $R_\ell$. Figure~\ref{recon_powers} shows the noiseless data
power spectrum, $C_\ell^{TT}=RP^\Phi R^\dag$, as well as the power spectrum $R^{(2)} P^\Phi_\ell R^{(2) \dag}$ expected from noiseless,
distance dependent data obtained with the estimator response, $d^{(2)}= R^{(2)}\Phi$. The $\ell-$blind spots are clearly recognizable,
which move from large scales at distances $r<r_\mathrm{LSS}$ to small scales at $r>r_\mathrm{LSS}$, where the amplitude of this power
spectrum gets maximal at $r=r_\mathrm{LSS}$. 

The numerical and computational effort to reconstruct one slice by one CPU amounts to roughly 45 minutes, which simultaneously represents 
the time for reconstructing the whole three-dimensional primordial scalar potential at full parallelization. In our numerical implementation
we used the \texttt{conjugate gradient} method to solve Eq.~(\ref{wiener}).

\section{Conclusion \& outlook}\label{sec:conclusion}
We have presented a reconstruction of the primordial scalar potential $\Phi$ with corresponding $1\sigma$-uncertainty from WMAP temperature
data. This has been achieved by setting up an inference 
approach that separates the whole inverse problem of reconstructing $\Phi$ into many independent ones,
each corresponding to the primordial scalar potential projected onto a sphere with specific comoving distance.
This way the reconstruction is done sphere by sphere until one obtains a thick shell of nested spheres around
the surface of last scattering. This results in a significant reduction of computational costs
(since the reconstruction equation (Wiener filter) parallelizes fully),
if only the small region around the last scattering surface is reconstructed, which is accessible
through CMB data. 

We did not include polarization information yet due to the suboptimal signal-to-noise ratios of the WMAP
polarization data. Hence we do not expect a huge improvement when additionally including WMAP
Stokes $Q$ and $U$ parameters into the Wiener filter equation. 
This, however, will definitely change when the polarization data of Planck will be available
in the near future. 
Once one uses simultaneously temperature and polarization data, the $\ell-$blind spots in the
reconstructions will disappear and with it the crumbliness of the maps. 
At this point it also might be more rewarding to apply the critical filer equations
to simultaneously obtain the power spectrum of the primordial scalar potential. 
 
\begin{acknowledgments}
We gratefully acknowledge Vanessa Boehm and Marco Selig for useful discussions and comments on the manuscript, as well as Eiichiro Komatsu for numerical
support concerning \texttt{gTfast}, to be found at \url{http://www.mpa-garching.mpg.de/~komatsu/CRL/nongaussianity/radiationtransferfunction/}. 
All calculations have been done using \textsc{NIFTy} \cite{2013AA...554A..26S} to be found at \url{http://www.mpa-garching.mpg.de/ift/nifty/},
in particular involving \texttt{HEALPix} \cite{0004-637X-622-2-759} to be found at \url{http://healpix.sourceforge.net/}. We also acknowledge
the support by the DFG Cluster of Excellence ``Origin and Structure of the Universe''. 
The calculations have been carried out on the computing facilities of the Computational Center for Particle and Astrophysics (C2PAP).
\end{acknowledgments}
\appendix
\section{Response projected onto the sphere of LSS}\label{app:response_der}
The data are given by
\begin{equation}
\begin{split}
 d_{\ell m} \equiv &~ M_{\ell m \atop \ell' m'}a^{\text{CMB}}_{\ell' m'} + n _{\ell m}= \left(R \Phi\right)_{\ell m} + n _{\ell m}\\
  =&~ M_{\ell m \atop \ell' m'} B_{\ell'} \frac{2}{\pi}\int dk ~k^2 \int dr~ r^2 \Phi_{\ell' m'}(r) g^T_{\ell'}(k)j_{\ell'}(kr) + n_{\ell m} .
\end{split}
\end{equation}
Considering Gaussian statistics for the primordial curvature perturbations, $\Phi$, the response is defined by
\begin{equation}
\label{response_3d}
 R \equiv \left\langle d \Phi^\dag \right\rangle_{(\Phi,d)} \left\langle \Phi \Phi^\dag \right\rangle^{-1}_{(\Phi,d)}.
\end{equation}
Instead of using the full three-dimensional response $R$, we introduce a two-dimensional response, $R^{(2)}$, which acts on the primordial potential
projected  onto the last scattering surface (LSS), $\Phi^{(2)} \equiv \Phi\left(r=r_\mathrm{LSS}\right) = \tilde{T}\Phi$, where $\tilde{T}$ denotes the projection operator:
\begin{equation}
R^{(2)} = \left\langle R\Phi (\tilde{T}\Phi)^\dag \right\rangle_{(\Phi,d)} \left\langle \tilde{T}\Phi (\tilde{T}\Phi)^\dag \right\rangle^{-1}_{(\Phi,d)} 
	= \left(R P^\Phi \tilde{T}^\dag\right)\left(\tilde{T} P^\Phi \tilde{T}^\dag \right)^{-1}.
\end{equation}
To derive the denominator at the distance of the LSS, we first transform it into position-space,
\begin{equation}
\begin{split}
 \left(\tilde{T} P^\Phi \tilde{T}^\dag \right)_{\mathbf{\hat{n}},\mathbf{\hat{n}'}} 
 =&~\int d^3\mathbf{x}\int d^3\mathbf{y} \delta\left(\mathbf{x} - r_{\text{LSS}}\mathbf{\hat{n}}\right)\delta\left(\mathbf{y} - r_{\text{LSS}}\mathbf{\hat{n}'}\right)\\
  &~\times \int \frac{d^3\mathbf{k}}{(2\pi)^3}\int \frac{d^3\mathbf{q}}{(2\pi)^3} (2\pi)^3 \delta(\mathbf{k}-\mathbf{q})  P^\Phi(k)	
      e^{-i \mathbf{k}\cdot\mathbf{x}} e^{ i \mathbf{q}\cdot\mathbf{y}}\\
 =&~\int \frac{d^3\mathbf{k}}{(2\pi)^3} P^\Phi(k) e^{-i r_{\text{LSS}} \mathbf{k}\cdot\mathbf{\hat{n}}} e^{i r_{\text{LSS}} \mathbf{k}\cdot\mathbf{\hat{n}'}}.
\end{split}
\end{equation}
Vectors are printed in bold for reasons of clarity and comprehensibility; unit vectors are denoted by $\hat{}$. Subsequently we use the Rayleigh expansion,
\begin{equation}
e^{i \mathbf{k}\cdot\mathbf{r}} = 4\pi \sum_{\ell=0}^{\infty} \sum_{m=-\ell}^{\ell} i^\ell j_\ell(kr) Y_\ell^{m *}(\mathbf{\hat{k}}) Y_\ell^{m}(\mathbf{\hat{r}}),  
\end{equation}
as well as the transformation rules
\begin{equation}
\begin{split}
f_{\ell m} \equiv&~ \oint d\mathbf{\hat{n}} ~Y_\ell^{m *}(\mathbf{\hat{n}})f(\mathbf{\hat{n}}),\\
		  f(\mathbf{\hat{n}}) =&~\sum_{\ell=0}^{\infty} \sum_{m=-\ell}^{\ell} f_{\ell m}Y_{\ell}^{m}(\mathbf{\hat{n}}),\\
\text{and}&~\oint d\mathbf{\hat{n}}~Y_{\ell}^{m}(\mathbf{\hat{n}}) Y_{\ell'}^{m' *}(\mathbf{\hat{n}})=\delta_{\ell \ell'}\delta_{m m'},
\end{split}
\end{equation}
to obtain the final corresponding expression in the spherical harmonic space, 
\begin{equation}
\begin{split}
\left( \tilde{T} P^\Phi \tilde{T}^\dag \right)_{\ell m \atop \ell' m'} =&~ \oint d\mathbf{\hat{n}} \oint d\mathbf{\hat{n}'}~ Y_\ell^{m}(\mathbf{\hat{n}}) Y_{\ell'}^{m' *}(\mathbf{\hat{n}'})
					    \int \frac{d^3\mathbf{k}}{(2\pi)^3} P^\Phi(k) \\
		  &~\times \sum_{\ell'' \ell'''\atop m'' m'''}(4\pi)^2 i^{\ell'''-\ell''} j_{\ell''}(kr_\text{LSS})j_{\ell'''}(kr_\text{LSS})
		    Y_{\ell''}^{m''}(\mathbf{\hat{k}}) Y_{\ell'''}^{m''' *}(\mathbf{\hat{k}}) Y_{\ell''}^{m'' *}(\mathbf{\hat{n}}) Y_{\ell'''}^{m'''}(\mathbf{\hat{n}'})\\
		  =&~\frac{2}{\pi}\int dk ~k^2 \oint d\mathbf{\hat{k}}~P^\Phi(k) i^{\ell'-\ell} j_{\ell}(kr_\text{LSS})j_{\ell'}(kr_\text{LSS})Y_{\ell'}^{m' *}(\mathbf{\hat{k}})Y_{\ell}^{m}(\mathbf{\hat{k}})\\
		  =&~\frac{2}{\pi}\int dk ~k^2 P^\Phi(k) j^2_{\ell}(kr_\text{LSS}) \delta_{\ell \ell'}\delta_{m m'} \equiv P_\ell^\Phi \delta_{\ell \ell'}\delta_{m m'}.
\end{split}
\end{equation}
$P_\ell^\Phi$ denotes the primordial power spectrum projected onto the sphere of LSS.

To determine the numerator we fist have to transform $P^\Phi \tilde{T}^\dag$ into the basis of spherical harmonics. Analogous to the calculation above we obtain
\begin{equation}
\left(P^\Phi \tilde{T}^\dag\right)_{\ell m \atop \ell' m'} (r) = \frac{2}{\pi}\int dk~ k^2 P^\Phi(k) j_{\ell}(kr_\text{LSS})j_{\ell}(kr) \delta_{\ell \ell'}\delta_{m m'}, 
\end{equation}
and thus
\begin{equation}
\begin{split}
\left(R P^\Phi \tilde{T}^\dag\right)_{\ell m \atop \ell' m'} =&~ M_{\ell m \atop \ell'' m''} B_{\ell''} \frac{2}{\pi}\\
&~\times \int dk~ k^2 \int dr ~r^2 \left\{ \frac{2}{\pi}\int dk' k'^2 P^\Phi(k') j_{\ell''}(k'r_\text{LSS})
j_{\ell''}(k'r) \right\}\\
&~\times g^T_{\ell''}(k)j_{\ell''}(kr) \delta_{\ell'' \ell'}\delta_{m'' m'}.
\end{split}
\end{equation}
Using the identity
\begin{equation}
\int_0^\infty dr~ r^2  j_{\ell}(kr)j_{\ell}(k'r) = \frac{\pi}{2}\frac{1}{k^2}\delta(k-k')
\end{equation}
finally yields
\begin{equation}
\left(R P^\Phi \tilde{T}^\dag\right)_{\ell m \atop \ell' m'} =M_{\ell m \atop \ell'' m''} B_{\ell''} \frac{2}{\pi}
\int dk ~k^2 P^\Phi(k) j_{\ell''}(kr_\text{LSS}) g^T_{\ell''}(k) \delta_{\ell'' \ell'}\delta_{m'' m'}.
\end{equation}

Putting the results together, the two-dimensional response is given by
\begin{equation}
\label{R2D}
 R_{\ell m \atop \ell' m'}^{(2)} =M_{\ell m \atop \ell'' m''} B_{\ell''} \frac{\int dk ~k^2 P^\Phi(k) j_{\ell''}(kr_\text{LSS}) 
 g^T_{\ell''}(k)}{\int dk ~k^2 P^\Phi(k) j^2_{\ell''}(kr_\text{LSS})}\delta_{\ell'' \ell'}\delta_{m'' m'}.
\end{equation}
The response for arbitrary comoving distances $r'$ can be obtained by replacing $r_\mathrm{LSS}$ by $r'$.


\begin{commenta}
\section{Data generation MAYBE ASLO RECONSTRUCTION ???}
Due to the projection of the primordial potential onto the sphere of LSS, we loose some power, 
\begin{equation}
\left\langle \left(R\Phi - R^{(2)}\Phi^{(2)}\right)\left(R\Phi - R^{(2)}\Phi^{(2)} \right)^\dag \right\rangle_{(\Phi)} = R^{\mathrm{mask}}_{\ell m} C_\ell R^{\mathrm{mask} \dag}_{\ell m} - R^{(2)}_{\ell m} P^\Phi_\ell R^{(2)}_{\ell m},
\end{equation}
with 
\begin{equation}
C_\ell =b^2_\ell \frac{2}{\pi} \int dk k^2 P^\Phi(k) g^2_\ell(k). 
\end{equation}

MAYBE: introduce a phenomenological scale parameter $\alpha$, defined by
\begin{equation}
 \alpha_\ell \equiv \sqrt{\frac{R^{\mathrm{mask}}_{\ell m} C_\ell R^{\mathrm{mask} \dag}_{\ell m}}{R_{\ell m}^{(2)} P^\Phi_\ell R^{(2)}_{\ell m}}} ~~\forall \ell: R^{(2)}_{\ell m} \neq 0,
\end{equation}
to correct for this power lost. Therefore, we introduce the phenomenological response 
\begin{equation}
R^*_{\ell m} \equiv \alpha_\ell  R^{(2)}_{\ell m}. 
\end{equation}
This way, we ensure that
\begin{equation}
\label{rescale}
\left\langle \left(R\Phi\right)\left(R\Phi\right)^\dag \right\rangle_{(\Phi)}=
\left\langle \left(R^*\Phi^{(2)}\right)\left(R^*\Phi^{(2)} \right)^\dag \right\rangle_{(\Phi^{(2)})} =R^{\mathrm{mask}}_{\ell m} C_\ell R^{\mathrm{mask} \dag}_{\ell m}\equiv C_{\ell m}^\mathrm{mask}. 
\end{equation}

\end{commenta}


\section{Wiener filter formula and uncertainty estimate in data space}\label{app:uncertain}
The Wiener filter in data space is defined by
\begin{equation}\label{app:wiener}
\begin{split}
 m^{(2)}_\text{w} \equiv &~\left\langle \Phi^{(2)} \right\rangle_{(\Phi|d)} = \tilde{T}\left\langle \Phi \right\rangle_{(\Phi|d)} = 
  \tilde{T}\left\langle \Phi d^\dag \right\rangle_{(\Phi,n)} \left\langle d d^\dag \right\rangle^{-1}_{(\Phi,n)} d = \tilde{T} P^\Phi R^\dag \left[R P^\Phi R^\dag
  + N \right]^{-1}d \\
  =&~ \tilde{T} P^\Phi R^\dag \left[\tilde{C}^{TT} + N \right]^{-1}d \stackrel{\text{Eq.~(\ref{R2D})}}{=} P_\ell^\Phi R^{(2) \dag} \left[\tilde{C}^{TT}  + N \right]^{-1}d.
\end{split}
\end{equation}

Formally, the corresponding posterior covariance matrix is constructed as
\begin{equation}  
\label{post_cov}
D = P_\ell^\Phi - P_\ell^\Phi R^{(2)\dagger}\left(\tilde{C}^{TT} + N\right)^{-1}  R^{(2)} P_\ell^\Phi.
\end{equation}
The square root of its position space diagonal would give us the 1$\sigma$ uncertainty map. However, as the operator is not directly accessible to us,
but is only defined as a sequence of linear functions, calculating the diagonal requires very expensive probing routines which need to evaluate the covariance matrix several thousand times before converging.
 
However, the covariance matrix becomes diagonal in spherical harmonic space under two conditions\footnote{Note that this procedure is only valid for a temperature-only analysis. Once
polarization data are included the 1$\sigma$ uncertainty must be calculated by the square root of the diagonal of Eq.~(\ref{post_cov}). }: We assume that there is no masking in the data and the noise covariance $N$ is a multiple of the identity. 
The noise covariance matrix for $TT$ data is already diagonal and dominated by white uncorrelated noise. 
So this approximation seems appropriate given the benefits in computational costs. The assumption that there is no masking is more drastic of course. 
We therefore construct our uncertainty map out of the limiting cases of having no masking and masking the whole sky. Both scenarios make the posterior covariance matrix diagonal in spherical harmonic space.
 
The constant approximation to the noise covariance is constructed as
\begin{equation}
 \tilde{N}_{\hat{n} \hat{n}'} = \frac{\mathrm{tr}\, N }{\mathrm{tr}\, \mathds{1} } \delta(\hat{n}-\hat{n}').
\end{equation}
The response with no mask is diagonal in spherical harmonic space,
\begin{equation}
 \tilde{R}_{{\ell m}\atop{\ell' m'}} = B_\ell R_\ell\, \delta_{\ell\ell'}\,\delta_{mm'},
\end{equation}
and the response with an all-sky mask is zero.
Therefore the covariance matrix is diagonal in either case.
Since a diagonal matrix in spherical harmonic space results in a constant diagonal in position space, we can exploit the invariance of the trace to get the position space diagonal of the covariance matrix,
\begin{equation}
 D_{\hat{n} \hat{n}} = \frac{\mathrm{tr}\,D}{4\pi},
\end{equation}
where the trace is easily calculated in spherical harmonic space, where $D$ is diagonal.
 
In a region that is fully masked and where the edges of the mask are further away than the correlation length of $\Phi$ the uncertainty approaches the limiting case of an all-sky mask.
In a region that is fully exposed and more than a correlation length away from a masked region the uncertainty approaches the limiting case of no mask.
We therefore combine the two cases into one map by setting the uncertainty to the ``all-sky masked'' value in regions which are masked and to the ``no mask'' value in regions which are not masked, i.e.
\begin{equation}
\label{sigma2}
  \sigma^2_{\hat{n}} = \begin{cases}
                       D_{\hat{n} \hat{n}}^{\mathrm{all\ mask}} & \mathrm{if}\ M_{\hat{n}\hat{n'}} = 0 \\
                       D_{\hat{n} \hat{n}}^{\mathrm{no\ mask}} & \mathrm{otherwise}. \\
                      \end{cases}
\end{equation}
The interpolation between these two regions is dictated by the prior covariance. It describes precisely how information is correlated between masked and unmasked regions.
Our final uncertainty map is therefore the result of a smoothing of $\sigma$ with the normalized square root of the prior covariance,
\begin{equation}
\label{sigma}
 \sigma_\mathrm{smooth} = \frac{1}{\mathcal{N}}\sqrt{P_\ell^\Phi} \sigma,
\end{equation}
where 
\begin{equation}
\mathcal{N} = \oint d\hat{n} d\hat{n}' \left(\sqrt{P_\ell^\Phi}\right)_{\hat{n}\hat{n}'}\, \delta(\hat{n}'). 
\end{equation}


\section{WMAP noise characterization}\label{app:noise}
The pixel noise level (in units mK) of a single map can be determined by $\sigma = \sigma_0 /\sqrt{N_\text{obs}}$, where $\sigma_0$ can be found 
at \url{http://lambda.gsfc.nasa.gov/product/map/dr5/skymap_info.cfm} and the effective number of observations $N_\text{obs}$, which can vary from pixel to pixel,
is stored in the FITS file of a map, see \url{http://lambda.gsfc.nasa.gov/product/map/dr4/skymap_file_format_info.cfm}.
Thus, the noise covariance matrix of a single map is given by
\begin{equation}
 N_{\mathbf{\hat{n}},\mathbf{\hat{n}'}} = \frac{\sigma_0^2}{N_\text{obs}(\mathbf{\hat{n}'})}\delta_{\mathbf{\hat{n}} \mathbf{\hat{n}'}}.
\end{equation}

Including polarization data, the noise covariance matrix in position space has to be generalized by

\begin{equation}
N^{-1} = 
\begin{pmatrix}
N^{TT}_{\mathrm{obs}}/\sigma_T^2 & 				0 & 				  0\\
0				& N^{QQ}_{\mathrm{obs}}/\sigma_P^2 & N^{QU}_{\mathrm{obs}}/\sigma_P^2\\
0				& N^{QU}_{\mathrm{obs}}/\sigma_P^2 & N^{UU}_{\mathrm{obs}}/\sigma_P^2
\end{pmatrix}
,
\end{equation}
where $\sigma_{T,P}$ is the respective noise level of temperature and polarization as given by WMAP. 

\bibliography{bibliography}

\providecommand{\noopsort}[1]{}\providecommand{\singleletter}[1]{#1}%
\providecommand{\href}[2]{#2}\begingroup\raggedright\begin{thebibliography}{10}

\bibitem{2009arXiv0907.5424B}
D.~{Baumann}, {\it {TASI Lectures on Inflation}},  {\em ArXiv e-prints} (July,
  2009) [\href{http://xxx.lanl.gov/abs/0907.5424}{{\tt arXiv:0907.5424}}].

\bibitem{2010AdAst2010E..71Y}
A.~P.~S. {Yadav} and B.~D. {Wandelt}, {\it {Primordial Non-Gaussianity in the
  Cosmic Microwave Background}},  {\em Advances in Astronomy} {\bf 2010} (2010)
  71, [\href{http://xxx.lanl.gov/abs/1006.0275}{{\tt arXiv:1006.0275}}].

\bibitem{2013MNRAS.432..894J}
J.~{Jasche} and B.~D. {Wandelt}, {\it {Bayesian physical reconstruction of
  initial conditions from large-scale structure surveys}},  {\em \mnras} {\bf
  432} (June, 2013) 894--913, [\href{http://xxx.lanl.gov/abs/1203.3639}{{\tt
  arXiv:1203.3639}}].

\bibitem{2014JCAP...07..012N}
S.~{Nurmi} and M.~S. {Sloth}, {\it {Constraints on gauge field production
  during inflation}},  {\em \jcap} {\bf 7} (July, 2014) 12,
  [\href{http://xxx.lanl.gov/abs/1312.4946}{{\tt arXiv:1312.4946}}].

\bibitem{2005PhRvD..71d3502M}
S.~{Matarrese}, S.~{Mollerach}, A.~{Notari}, and A.~{Riotto}, {\it {Large-scale
  magnetic fields from density perturbations}},  {\em \prd} {\bf 71} (Feb.,
  2005) 043502, [\href{http://xxx.lanl.gov/abs/astro-ph/0410687}{{\tt
  astro-ph/0410687}}].

\bibitem{2013arXiv1303.5084P}
{Planck Collaboration}, P.~A.~R. {Ade}, N.~{Aghanim}, C.~{Armitage-Caplan},
  M.~{Arnaud}, M.~{Ashdown}, F.~{Atrio-Barandela}, J.~{Aumont},
  C.~{Baccigalupi}, A.~J. {Banday}, and et~al., {\it {Planck 2013 Results.
  XXIV. Constraints on primordial non-Gaussianity}},  {\em ArXiv e-prints}
  (Mar., 2013) [\href{http://xxx.lanl.gov/abs/1303.5084}{{\tt
  arXiv:1303.5084}}].

\bibitem{paper1}
S.~{Dorn}, N.~{Oppermann}, R.~{Khatri}, M.~{Selig}, and T.~A. {En{\ss}lin},
  {\it {Fast and precise way to calculate the posterior for the local
  non-Gaussianity parameter f$_{nl}$ from cosmic microwave background
  observations}},  {\em \prd} {\bf 88} (Nov., 2013) 103516,
  [\href{http://xxx.lanl.gov/abs/1307.3884}{{\tt arXiv:1307.3884}}].

\bibitem{2014JCAP...06..048D}
S.~{Dorn}, E.~{Ramirez}, K.~E. {Kunze}, S.~{Hofmann}, and T.~A. {En{\ss}lin},
  {\it {Generic inference of inflation models by non-Gaussianity and primordial
  power spectrum reconstruction}},  {\em \jcap} {\bf 6} (June, 2014) 48,
  [\href{http://xxx.lanl.gov/abs/1403.5067}{{\tt arXiv:1403.5067}}].

\bibitem{2013ApJS..208...20B}
C.~L. {Bennett}, D.~{Larson}, J.~L. {Weiland}, N.~{Jarosik}, G.~{Hinshaw},
  N.~{Odegard}, K.~M. {Smith}, R.~S. {Hill}, B.~{Gold}, M.~{Halpern},
  E.~{Komatsu}, M.~R. {Nolta}, L.~{Page}, D.~N. {Spergel}, E.~{Wollack},
  J.~{Dunkley}, A.~{Kogut}, M.~{Limon}, S.~S. {Meyer}, G.~S. {Tucker}, and
  E.~L. {Wright}, {\it {Nine-year Wilkinson Microwave Anisotropy Probe (WMAP)
  Observations: Final Maps and Results}},  {\em \apjs} {\bf 208} (Oct., 2013)
  20, [\href{http://xxx.lanl.gov/abs/1212.5225}{{\tt arXiv:1212.5225}}].

\bibitem{2009PhRvD..80j5005E}
T.~A. {En{\ss}lin}, M.~{Frommert}, and F.~S. {Kitaura}, {\it {Information field
  theory for cosmological perturbation reconstruction and nonlinear signal
  analysis}},  {\em \prd} {\bf 80} (Nov., 2009) 105005,
  [\href{http://xxx.lanl.gov/abs/0806.3474}{{\tt arXiv:0806.3474}}].

\bibitem{2005ApJ...634...14K}
E.~{Komatsu}, D.~N. {Spergel}, and B.~D. {Wandelt}, {\it {Measuring Primordial
  Non-Gaussianity in the Cosmic Microwave Background}},  {\em \apj} {\bf 634}
  (Nov., 2005) 14--19, [\href{http://xxx.lanl.gov/abs/astro-ph/}{{\tt
  astro-ph/}}].

\bibitem{wiener1964time}
N.~Wiener, {\em Extrapolation, Interpolation, and Smoothing of Stationary Time
  Series}.
\newblock New York: Wiley, 1949.

\bibitem{2005PhRvD..71l3004Y}
A.~P. {Yadav} and B.~D. {Wandelt}, {\it {CMB tomography: Reconstruction of
  adiabatic primordial scalar potential using temperature and polarization
  maps}},  {\em \prd} {\bf 71} (June, 2005) 123004,
  [\href{http://xxx.lanl.gov/abs/astro-ph/0505386}{{\tt astro-ph/0505386}}].

\bibitem{2014A&A...571A..16P}
{Planck Collaboration}, P.~A.~R. {Ade}, N.~{Aghanim}, C.~{Armitage-Caplan},
  M.~{Arnaud}, M.~{Ashdown}, F.~{Atrio-Barandela}, J.~{Aumont},
  C.~{Baccigalupi}, A.~J. {Banday}, and et~al., {\it {Planck 2013 results. XVI.
  Cosmological parameters}},  {\em \aap} {\bf 571} (Nov., 2014) A16,
  [\href{http://xxx.lanl.gov/abs/1303.5076}{{\tt arXiv:1303.5076}}].

\bibitem{1997PhRvD..55.1830Z}
M.~{Zaldarriaga} and U.~{Seljak}, {\it {All-sky analysis of polarization in the
  microwave background}},  {\em \prd} {\bf 55} (Feb., 1997) 1830--1840,
  [\href{http://xxx.lanl.gov/abs/astro-ph/9609170}{{\tt astro-ph/9609170}}].

\bibitem{2001PhRvD..63f3002K}
E.~{Komatsu} and D.~N. {Spergel}, {\it {Acoustic signatures in the primary
  microwave background bispectrum}},  {\em \prd} {\bf 63} (Mar., 2001) 063002,
  [\href{http://xxx.lanl.gov/abs/astro-ph/}{{\tt astro-ph/}}].

\bibitem{2014A&A...571A..22P}
{Planck Collaboration}, P.~A.~R. {Ade}, N.~{Aghanim}, C.~{Armitage-Caplan},
  M.~{Arnaud}, M.~{Ashdown}, F.~{Atrio-Barandela}, J.~{Aumont},
  C.~{Baccigalupi}, A.~J. {Banday}, and et~al., {\it {Planck 2013 results.
  XXII. Constraints on inflation}},  {\em \aap} {\bf 571} (Nov., 2014) A22,
  [\href{http://xxx.lanl.gov/abs/1303.5082}{{\tt arXiv:1303.5082}}].

\bibitem{2014A&A...566A..77P}
P.~{Paykari}, F.~{Lanusse}, J.-L. {Starck}, F.~{Sureau}, and J.~{Bobin}, {\it
  {PRISM: Sparse recovery of the primordial power spectrum}},  {\em \aap} {\bf
  566} (June, 2014) A77, [\href{http://xxx.lanl.gov/abs/1402.1983}{{\tt
  arXiv:1402.1983}}].

\bibitem{2011PhRvD..83j5014E}
T.~A. {En{\ss}lin} and M.~{Frommert}, {\it {Reconstruction of signals with
  unknown spectra in information field theory with parameter uncertainty}},
  {\em \prd} {\bf 83} (May, 2011) 105014,
  [\href{http://xxx.lanl.gov/abs/1002.2928}{{\tt arXiv:1002.2928}}].

\bibitem{2012arXiv1210.6866O}
N.~{Oppermann}, M.~{Selig}, M.~R. {Bell}, and T.~A. {En{\ss}lin}, {\it
  {Reconstruction of Gaussian and log-normal fields with spectral smoothness}},
   {\em ArXiv e-prints} (Oct., 2012)
  [\href{http://xxx.lanl.gov/abs/1210.6866}{{\tt arXiv:1210.6866}}].

\bibitem{2013ApJS..208...19H}
G.~{Hinshaw}, D.~{Larson}, E.~{Komatsu}, D.~N. {Spergel}, C.~L. {Bennett},
  J.~{Dunkley}, M.~R. {Nolta}, M.~{Halpern}, R.~S. {Hill}, N.~{Odegard},
  L.~{Page}, K.~M. {Smith}, J.~L. {Weiland}, B.~{Gold}, N.~{Jarosik},
  A.~{Kogut}, M.~{Limon}, S.~S. {Meyer}, G.~S. {Tucker}, E.~{Wollack}, and
  E.~L. {Wright}, {\it {Nine-year Wilkinson Microwave Anisotropy Probe (WMAP)
  Observations: Cosmological Parameter Results}},  {\em \apjs} {\bf 208} (Oct.,
  2013) 19, [\href{http://xxx.lanl.gov/abs/1212.5226}{{\tt arXiv:1212.5226}}].

\bibitem{1996ApJ...469..437S}
U.~{Seljak} and M.~{Zaldarriaga}, {\it {A Line-of-Sight Integration Approach to
  Cosmic Microwave Background Anisotropies}},  {\em \apj} {\bf 469} (Oct.,
  1996) 437, [\href{http://xxx.lanl.gov/abs/astro-ph/}{{\tt astro-ph/}}].

\bibitem{2013AA...554A..26S}
M.~{Selig}, M.~R. {Bell}, H.~{Junklewitz}, N.~{Oppermann}, M.~{Reinecke},
  M.~{Greiner}, C.~{Pachajoa}, and T.~A. {En{\ss}lin}, {\it {NIFTY - Numerical
  Information Field Theory. A versatile PYTHON library for signal inference}},
  {\em \aap} {\bf 554} (June, 2013) A26,
  [\href{http://xxx.lanl.gov/abs/1301.4499}{{\tt arXiv:1301.4499}}].

\bibitem{0004-637X-622-2-759}
K.~M. G{\`{o}}rski, E.~Hivon, A.~J. Banday, B.~D. Wandelt, F.~K. Hansen,
  M.~Reinecke, and M.~Bartelmann, {\it Healpix: A framework for high-resolution
  discretization and fast analysis of data distributed on the sphere},  {\em
  The Astrophysical Journal} {\bf 622} (2005), no.~2 759.

\end{thebibliography}\endgroup
\end{document}